\documentclass[preprint]{aastex}

\begin{document}
                                         
\title{Complementary constraints from FR IIb radio galaxies and X-ray gas mass fractions in clusters on non-standard cosmological models}
             
%short titel {FRIIb/X-ray cluster gas fraction constraints}

\author{Dirk Puetzfeld, Martin Pohl}
\affil{Department of Physics and Astronomy, Iowa State University, Ames, IA 50011, USA}
\email{dpuetz@iastate.edu, mkp@iastate.edu}
\and
\author{Zong-Hong Zhu}
\affil{Department of Astronomy, Beijing Normal University, Beijing 100875, CHINA}
\email{zhuzh@bnu.edu.cn}

\begin{abstract}
We use recent measurements of the dimensionless coordinate distances from Fanaroff-Riley Type IIb radio galaxies and the X-ray gas mass fractions in clusters to constrain the parameters of a non-standard cosmological model. This work complements our recent analysis of the SN Ia data within a non-Riemannian cosmological model. We use two independent data sets to constrain the new density parameter $\Omega_\psi$, which is related to the non-Riemannian structure of the underlying spacetime and supplements the field equations that are very similar to the usual Friedmann equations of general relativity. Thereby we place an upper limit on the presence of non-Riemannian quantities in the late stages of the universe. The numerical results of this work also apply to several anisotropic cosmological models which, on the level of the field equations, exhibit a similar scaling behavior of the density parameters like our non-Riemannian model.   
\end{abstract}

\maketitle

\section{Introduction\label{Introduction_section}}
From an observational standpoint, cosmology seems to be in a rather good shape. With several independent cosmological tests at hand the so-called standard model of cosmology \citep{KolbTurner,PadmanabhanAstro3} has emerged and passed most of these tests. Today cosmology is based on observations related to the global expansion of the universe, the cosmic microwave background, primordial nucleosynthesis, and the formation of structure in the universe. From a theoretical standpoint the standard cosmological model relies on the theory of general relativity or, to be more precise, on a homogeneous and isotropic model, the so-called FLRW model (named after Friedmann, Lema\^{\i}tre, Robertson, and Walker). While one of the main benefits of the FLRW model is given by its simplicity, recent observations of type Ia supernovae \citep{Hamuy1996,Perlmutter2,Perlmutter,Garnavich1,Schmidt,Riess2,Riess1,Barris,Tonry} made clear that we know little about the dominating energy density component of the universe, which enters the general relativistic description in the form of a cosmological constant and is nowadays termed dark energy. Several mechanisms have been proposed during the last years in order to remove the need for the usual cosmological constant. Among them are evolving scalar fields \citep{Ratra,Wetterich}, a time varying cosmological ``constant'' \citep{Ozer,Vishwakarma3}, $k$-essence \citep{Chiba,Armendariz}, a phantom energy \citep{Caldwell}, the Chaplygin gas \citep{Chaplygin0,Chaplygin1,Chaplygin2,Chaplygin3,Chaplygin4,Chaplygin5,Chaplygin7,Chaplygin9}, a modification of the FLRW equation termed ``Cardassian expansion'' \citep{FreeseLewis,Zhu2002,Zhu2003}, and the embedding of our universe in a higher dimensional bulk spacetime \citep{Randall1,Randall2,Deffayet}.  

In \citet{PuetzfeldChen} we constrained the parameters of a non-standard cosmological model with the help of recent SN Ia data sets of \citet{Wang} and \citet{Tonry}. The model investigated therein is based on a non-Riemannian spacetime, the so-called Weyl-Cartan spacetime, and was analyzed by several groups during the last few years \citep{PuetzfeldChen, Puetzfeld1, Puetzfeld2, Obukhov, Babourova}. For an overview of the developments in non-Riemannian cosmology see \citet{Puetzfeld2004}. Our main aim in \citet{PuetzfeldChen} was to place an upper limit on the new density parameter $\Omega_\psi$ using the latest SNe data. This parameter is linked to the presence of non-Riemannian field strengths which might play an important role with respect to the observed accelerated expansion of the universe. In this work we pin down the parameters of our model by means of the recently released FR IIb radio galaxy data set of \citet{Daly2003} and the data on X-ray gas mass fractions in clusters as provided in \citet{Allen2002,Allen2003,Allen2004}.   

The reason to consider FR IIb radio galaxies as a distance measure is twofold. At the moment the redshift range covered by the SNe ranges up to $z=1.7$ with only very few data points above $z=1$\footnote{This situation has changed a bit while this work was nearly finished, cf.\ \citet{Riess2004}.}. The FR IIb radio galaxy data set of \citet{Daly2003} contains eight data points with $z>1$ and thereby significantly enhances the available sample at high redshifts. Apart from the enhanced redshift range coverage the FR IIb radio galaxies provide an independent cosmological test since one deals with different objects. To clearly separate the parameter estimates from FR IIb galaxies and SN Ia we do {\it not} use a combined set in our fitting procedure. One of tasks in this paper is to determine if the cosmological parameters extracted from the FR IIb data set are compatible with the results from the SN Ia obtained in \cite{PuetzfeldChen}. The same arguments, apart from the coverage of high redshifts, also hold for the X-ray gas mass fraction measurements in clusters \citep{Allen2002,Allen2003,Allen2004}.

The plan of the paper is as follows. In section \ref{Field_eq_in_ext_model_section} we provide a short derivation of the distance redshift relations which are needed to perform the fits. Thereafter we introduce the different data sets in section \ref{Numerical_results_section} and briefly describe our fitting method. In section \ref{CONCLUSION_section} we compare our findings with the results of \citet{PuetzfeldChen} and draw our final conclusion. Readers who are not familiar with the model under consideration might want to consider \citet{Obukhov, Puetzfeld1, Babourova}, and \citet{PuetzfeldChen} first. Appendix A of \citet{PuetzfeldChen} contains a brief introduction to the field equations and geometrical quantities of metric-affine gravity (MAG) and to the so-called triplet ansatz of MAG. Readers who are not familiar with MAG might also want to consult \citet{PhysRep} for a comprehensive review. Appendix \ref{APP_data_sets} contains the two data sets of \citet{Daly2003} and \citet{Allen2004} that we use in our fitting procedure.
    
\section{Basic model equations}

\paragraph{Field equations\label{Field_eq_in_ext_model_section}}

By making a triplet ansatz for torsion and nonmetricity and by using the usual Robertson-Walker line element
\begin{equation}
ds^2=-dt^2+S(t)^2\left(\frac{dr^2}{1-kr^2}+r^2 d \theta^2 +r^2 \sin^2(\theta)d\phi^2\right), \label{RW_line_element}
\end{equation}
the general field equations of the model considered in \citet{PuetzfeldChen} reduced to the set 
\begin{eqnarray}
\frac{\dot{S}^{2}}{S^{2}}+\frac{k}{S^{2}} &=&\frac{\kappa }{3}\left[ \mu + \frac{\kappa }{48a_{0}}\left( 1-\frac{3a_{0}}{b_{4}}\right) \frac{\psi ^{2}}{S^{6}}\right] ,  \label{EXT_obukhov_triplet1} \\
2\frac{\ddot{S}}{S}+\frac{\dot{S}^{2}}{S^{2}}+\frac{k}{S^{2}} &=&-\kappa \left[ p+\frac{\kappa }{48a_{0}}\left( 1-\frac{3a_{0}}{b_{4}}\right) \frac{\psi ^{2}}{S^{6}}\right].  \label{EXT_obukhov_triplet2}
\end{eqnarray}
Here $a_{0}$ and $b_{4}$ are the coupling constants from the Lagrangian in eq.\ (1) of \citet{PuetzfeldChen} and $\psi $ denotes an integration constant entering the solution for the Weyl 1-form $Q$ which is given by\footnote{Note that we changed some of the variable names of \citet{Obukhov}, in which this model was firstly investigated, in order to match our notation in \citet{Puetzfeld1,Puetzfeld2}.} 
\begin{equation}
Q=-\frac{\kappa \psi }{8b_{4}}S^{-3}dt.  \label{EXT_obukhov_triplet_1_form}
\end{equation}
As one can easily see the field equations are the usual Friedmann equations, without a cosmological constant, with an additional contribution to the energy and pressure from the dilation current. The non-Riemannian quantities in this model, i.e.\ torsion and nonmetricity, {\it die out} as the universe expands. 

On the level of the field equations this model proves to be compatible with the one proposed in \citet{Puetzfeld1} if we make the choice $a_6=-a_4$ for the coupling constants in the Lagrangian in equation (1) of \citet{Puetzfeld1}. Additionally, one can show that also the model of \citet{Babourova} yields the same set of field equations, if one performs the transformations listed in eq.\ (7) of \cite{PuetzfeldChen}.

In the next section we outline the derivation of several distance notions within this model. In contrast to the original model in \cite{Obukhov} we explicitly allow for a cosmological constant, which corresponds to an additional term $-\lambda/3$ on the lhs of (\ref{EXT_obukhov_triplet1}) and an extra $-\lambda$ on the lhs of (\ref{EXT_obukhov_triplet2}).

\paragraph{Distance relations\label{SEC_angular_diameter_distance}}

By defining a new constant $\upsilon :=\frac{\kappa ^{2}}{144 a_0}\left( 1-\frac{3a_{0}}{b_{4}}\right) $ we can rewrite (\ref{EXT_obukhov_triplet1}) according to 
\begin{equation}
1+\frac{k}{S^{2}H^{2}}-\frac{\lambda}{3}=\frac{\kappa }{3 H^2}\mu +\upsilon \frac{\psi ^{2}}{S^{6} H^2}\quad \Rightarrow \quad \Omega _{k}+\Omega_{\lambda}+\Omega _{w}+\Omega _{\psi }=1.
\label{EXT_obuk_relation_between_density_parameters}
\end{equation}
Here we introduced the following density parameters $\Omega _{k}:=-\frac{k}{H^{2}S^{2}}$, $\Omega_\lambda=\frac{\lambda}{3 H^2}$, $\Omega _{w}:=\frac{\kappa }{3 H^2}\mu $, $\Omega _{\psi}:=\upsilon \frac{\psi ^{2}}{S^{6} H^2}$ in the last step. We use the index $w$ since we did not fix the underlying equation of state, $p=w\mu$. It is interesting to note that the new density parameter $\Omega_{\psi}$ redshifts with $z^6$, a behavior which is also known from anisotropic models, e.g.\ see \citet{KamionTurn,Khalatnikov}. Denoting present-day values of quantities by an index ``0'' we can rewrite the Hubble rate in terms of the density parameters and the redshift:
\begin{eqnarray}
H^{2}=\frac{\kappa }{3}\mu -\frac{k}{S^{2}}+\frac{\lambda}{3}+\upsilon \frac{\psi ^{2}}{S^{6}} = H_{0}^{2}\left[ \Omega _{w0}\left( 1+z\right) ^{3\left( 1+w\right)}+\Omega _{k0}\left( 1+z\right) ^{2}+\Omega_{\lambda 0}+\Omega _{\psi 0}\left( 1+z\right) ^{6}\right]   \nonumber \\
\stackrel{(\ref{EXT_obuk_relation_between_density_parameters})}{=}H_{0}^{2}\left( 1+z\right) ^{2}\left\{ \phantom{\frac{}{}}\Omega _{w0}\left[ \left(1+z\right) ^{1+3w}-1\right] +\Omega _{\lambda 0}\left[ \left( 1+z\right) ^{-2}-1\right] +\Omega _{\psi 0}\left[ \left( 1+z\right) ^{4}-1\right] +1  \phantom{\frac{}{}} \right\}. \label{EXT_obuk_Hubble_rate_with_density}
\end{eqnarray}
Hence the luminosity distance within this model becomes
\begin{equation}
d_{\texttt{\tiny \rm \rm lum}}=S_{0}\,\,\left( 1+z\right) \,\Theta \left[ \left(H_{0}S_{0}\right) ^{-1}\int_{0}^{z}F\left[ \tilde{z}\right] d\tilde{z}\right]
.  \label{EXT_obuk_general_luminosity}
\end{equation}
With $F[\tilde{z}]:=H_0/H$ and the function in front of the integral is given by 
\begin{equation}
\Theta \lbrack x]:=\left\{\begin{tabular}{lll}
$\sin \left( x\right) $ &  & $k=+1$ \\ 
$x$ & for & $k=0$ \\ 
$\sin $h$\left( x\right) $ &  & $k=-1$
\end{tabular}
\right. .\label{theta_definition}
\end{equation}
If we make use of the definition of the density parameter $\Omega _{k}$ we end up with
\begin{eqnarray}
d_{\texttt{\tiny \rm \rm lum}}\left( z;H_{0},\Omega _{w0},\Omega _{\lambda 0},\Omega _{\psi 0},w\right) = (1+z)^2 \, d_{\texttt{\tiny \rm \rm ang}}\left( z;H_{0},\Omega _{w0},\Omega _{\lambda 0},\Omega _{\psi 0},w\right) \nonumber \\ = \,\frac{\left( 1+z\right) }{H_{0}\sqrt{\left| 1-\Omega _{w0}-\Omega _{\lambda 0}-\Omega _{\psi0}\right| }}\,\Theta \left[ \sqrt{\left| 1-\Omega _{w0}-\Omega _{\lambda 0}-\Omega _{\psi0}\right| }\int_{0}^{z}F\left[ \tilde{z}\right] d\tilde{z}\right] .
\label{EXT_obuk_luminosity_with_omega_and_z}
\end{eqnarray}
We define the dimensionless coordinate distance $y:= d_{\texttt{\tiny \rm lum}} \, H_0 / (1+z)$, which is now given by
\begin{equation}
y\left( z;\Omega _{w0},\Omega _{\lambda 0},\Omega _{\psi 0},w\right)=\frac{\Theta \left[ \sqrt{\left| 1-\Omega _{w0}-\Omega _{\lambda 0}-\Omega _{\psi0}\right| }\int_{0}^{z}H_0/H\left( \tilde{z};H_0, \Omega _{w0},\Omega _{\lambda 0},\Omega _{\psi 0},w\right) d\tilde{z}\right]}{\sqrt{\left| 1-\Omega _{w0}-\Omega _{\lambda 0}-\Omega _{\psi0}\right| }}.
\label{EXT_dimensionless_coordinate_distance}
\end{equation}
The magnitude-redshift relation reads
\begin{equation}
m\left( z,H_{0},\Omega _{w0},\Omega _{\lambda 0},\Omega _{\psi 0},w,M\right) =M+5\log \left(\frac{d_{\texttt{\tiny \rm \rm lum}}}{{\rm length}}\right)+25.
\label{EXT_obuk_magnitude_redshift}
\end{equation}
In the case of a model with pressureless matter, radiation, and a contribution from the cosmological constant the dimensionless coordinate distance from equation (\ref{EXT_dimensionless_coordinate_distance}) is explicitly given by
\begin{eqnarray}
y\left( z;\Omega _{{\rm m}0},\Omega _{\lambda 0},\Omega _{\psi 0},\Omega _{{\rm r}0}\right)={\left| 1-\Omega _{{\rm m}0}-\Omega _{\lambda 0}-\Omega _{\psi0}-\Omega _{{\rm r}0}\right|^{-\frac{1}{2}}}\, \Theta \left[ \sqrt{\left| 1-\Omega _{{\rm m}0}-\Omega _{\lambda 0}-\Omega _{\psi0}-\Omega _{{\rm r}0}\right| }\right. \nonumber \\
\left. \int_{0}^{z} \left( 1+\tilde{z}\right) ^{-1}\left\{ \phantom{\frac{}{}}\Omega _{{\rm m}0} \tilde{z} + \Omega _{{\rm r}0} \left[ \left(1+\tilde{z}\right)^{2}-1\right] +\Omega _{\lambda 0}\left[ \left( 1+\tilde{z}\right) ^{-2}-1\right] +\Omega _{\psi 0}\left[ \left( 1+\tilde{z}\right) ^{4}-1\right] +1  \phantom{\frac{}{}} \right\}^{-\frac{1}{2}} d\tilde{z} \right].
\label{EXT_dimensionless_coordinate_distance_explicit}
\end{eqnarray}
We make use of relation (\ref{EXT_dimensionless_coordinate_distance_explicit}) in the next section where we perform fits to the FR IIb data set of \citet{Daly2003}, see also \citet{Daly1994,Guerra2000,Daly2002} and references therein.
\paragraph{Gas mass fraction}In addition to the FR IIb radio galaxies we also make use of the X-ray gas mass fraction measurements in dynamically relaxed clusters. Originally this method was described in \citet{Sasaki1996} and \citet{Pen1997}. Recent measurements of the gas mass fraction were performed by \citet{Allen2002,Allen2003,Allen2004}. Following their work we can test a given cosmological model by using 
\begin{equation}
f_{\rm gas}\left(z;H_0,b,\Omega _{{\rm m}0}, \Omega_{{\rm b}0}, \Omega _{\lambda 0},\Omega _{\psi 0},\Omega _{{\rm r}0}\right) = \frac{b \, \Omega_{{\rm b}0}}{\left(1+0.19\,\sqrt{h}\right) \Omega_{{\rm m}0}} \left[\frac{h}{0.5}\frac{d_{\texttt{\tiny \rm \rm ang}}^{\texttt{\tiny \rm \rm CDM}} \left( z;H_{0}\right)}{d_{\texttt{\tiny \rm \rm ang}}\left( z;H_{0},\Omega _{{\rm m}0},\Omega _{\lambda 0},\Omega _{\psi 0},\Omega _{{\rm r}0}\right)}\right]^{\frac{3}{2}}.
\label{gas_mass_fraction_Allen_et_al}
\end{equation}
Here $b$ is a bias factor motivated by gas dynamical simulations that takes account for the fact that the baryon fraction in clusters seems to be lower than for the universe as a whole, cf.\ \citet{Cen,Eke1998,Bialek} and references therein. The parameter $h$ stems from the parametrization $H_0=100 \, h \, {\rm km \, s}^{-1}{\rm Mpc}^{-1}$, and $\Omega_{{\rm b}0}$ corresponds to the baryonic matter content of the universe in terms of the critical density. The angular diameter distance for the standard CDM model with $\Omega_{{\rm m}0}=1$ is denoted by $d_{\texttt{\tiny \rm \rm ang}}^{\texttt{\tiny \rm \rm CDM}}$, it can be easily obtained by setting all other density parameters in (\ref{EXT_obuk_luminosity_with_omega_and_z}) to zero. The prefactor $\left(1+0.19 \sqrt{h} \right)$ stems from a conversion of the X-ray gas mass into the total baryonic mass, cf.\ \citet{Fukugita1998}.

\section{Numerical results\label{Numerical_results_section}}

\paragraph{Data sets\label{SUB_data_sets}}

\begin{figure}
\includegraphics[angle=-90,scale=0.55]{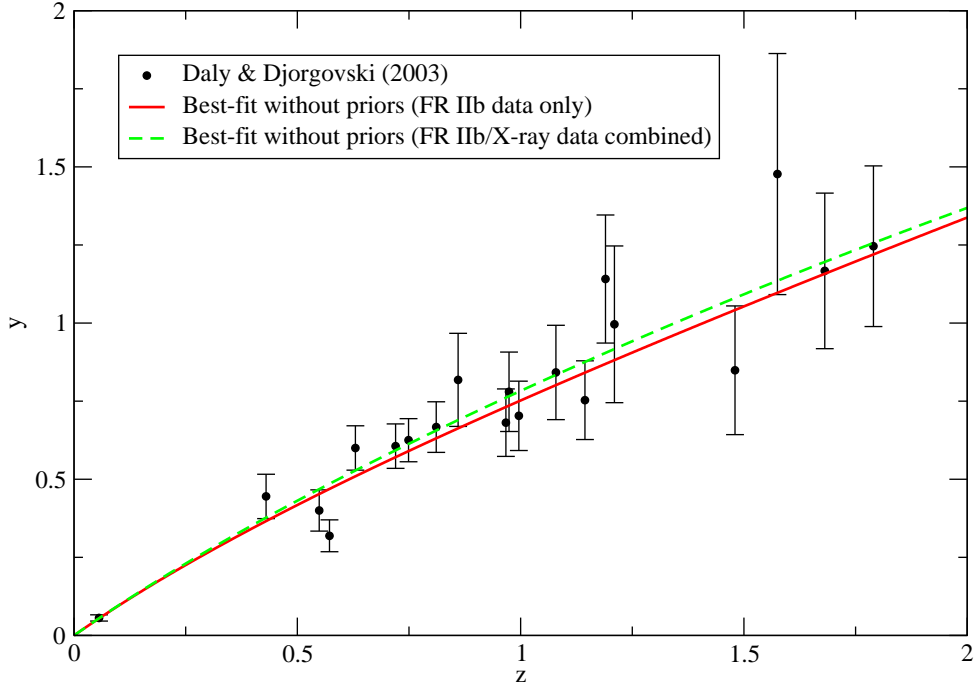}
\caption{Data set containing 20 FR IIb radio galaxies as compiled by \citet{Daly2003}, cf.\ table \ref{TAB_radio_galaxy_data}. The plots for the dimensionless coordinate distance correspond to the best-fit parameters given in table \ref{TAB_FRII_fit_results} and table \ref{TAB_combined_results}.}
\label{FIG_Daly_data_set}
\end{figure}

We have plotted the data set of \citet{Daly2003} consisting of the dimensionless distance parameter for 20 FR IIb radio galaxies in figure \ref{FIG_Daly_data_set}. To be as clear as possible we also listed the corresponding numerical values in the table \ref{TAB_radio_galaxy_data} of the appendix.
The combined data set from \citet{Allen2004} for the gas mass fraction of 26 clusters are displayed in figure \ref{FIG_Allen_data_set} and table \ref{TAB_x_ray_data_new} of the appendix. In figure \ref{FIG_Allen_data_set} we also display the 9 older data points from \citet{Allen2002,Allen2003}. Note that in our fits we always make use of the newer data set which comprises 26 clusters. 

\placefigure{FIG_Allen_data_set}
\placefigure{FIG_Daly_data_set}

\begin{figure}
\includegraphics[angle=-90,scale=0.55]{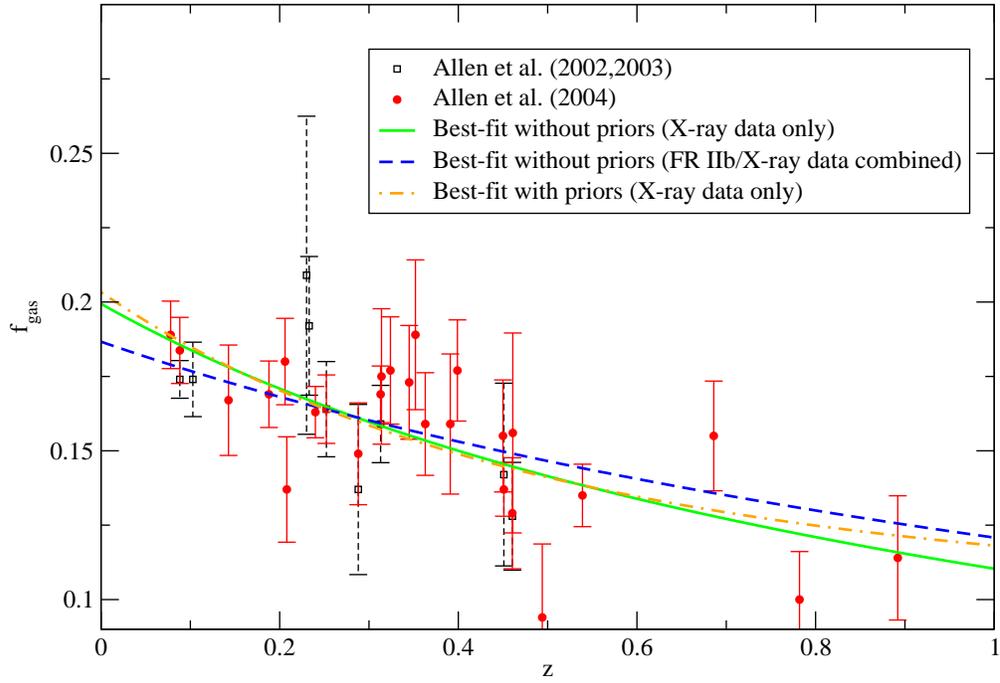}
\caption{Data set of the X-ray gas mass fractions in clusters of galaxies as given by \citet{Allen2002,Allen2003,Allen2004}. We always use the newer data set with 26 data points in our fit procedure, cf.\ table \ref{TAB_x_ray_data_new}. The plots of $f_{\rm gas}$ correspond to the best-fit parameters given in tables \ref{TAB_x_ray_fit_results}--\ref{TAB_x_ray_with_assumptions_results}.}
\label{FIG_Allen_data_set}
\end{figure}

\paragraph{Fitting method\label{SUB_fitting_method}}

In order to determine the best-fit parameters from the data set of \citet{Daly2003} we minimize
\begin{equation}
\chi^2\left(\Omega _{{\rm m}0},\Omega _{\lambda 0},\Omega _{\psi 0},\Omega _{{\rm r}0}\right):=\sum\limits_{i=1}^{20}\frac{\left[y_i^{\rm theory}\left(z_i ;  \Omega _{{\rm m}0},\Omega _{\lambda 0},\Omega _{\psi 0},\Omega _{{\rm r}0}\right)-y_i^{\rm measured}\right]^2}{\sigma^2_{y_i}}, \label{chi_square_radio}
\end{equation}
where $y_i^{\rm theory}\left(z_i ;  \Omega _{{\rm m}0},\Omega _{\lambda 0},\Omega _{\psi 0},\Omega _{{\rm r}0}\right)$ denotes the dimensionless coordinate distance from equation (\ref{EXT_dimensionless_coordinate_distance_explicit}). Note that we do not use any priors in this relation.

For the gas mass fractions we use the fitting formula provided by \citet{Allen2004}, namely 
\begin{eqnarray}
\chi^2\left(H_0, b, \Omega _{{\rm m}0}, \Omega_{{\rm b}0}, \Omega _{\lambda 0},\Omega _{\psi 0},\Omega _{{\rm r}0}\right)&:=\sum\limits_{i=1}^{26}\frac{\left[f_{{\rm gas},i}^{\rm theory}\left(z;H_0, b, \Omega _{{\rm m}0}, \Omega_{{\rm b}0}, \Omega _{\lambda 0},\Omega _{\psi 0},\Omega _{{\rm r}0}\right)-f_{{\rm gas},i}^{\rm measured}\right]^2}{\sigma^2_{f_{{\rm gas},i}}} \nonumber \\
&+\left(\frac{\Omega_{{\rm b}0} \, h^2 - 0.0214}{0.002} \right)^2 + \left( \frac{h-0.72}{0.08}\right)^2 + \left( \frac{b-0.824}{0.089}\right)^2 . 
 \label{chi_gas_frac}
\end{eqnarray}
The numerical value for the baryonic density parameter $\Omega_{{\rm b}0}\,h^2=0.0214\pm 0.002$, which was determined by the D/H ratio toward Q1243+3047 \citep{Kirkman2003}, is slightly lower than the WMAP result $\Omega_{{\rm b}0}\,h^2=0.024\pm 0.001$ but in good agreement with the combined WMAP+ACBAR+ CBI+2dFGRS determination of $\Omega_{{\rm b}0}\,h^2=0.0224\pm 0.0009$ as given in \citet{Spergel2003}. In the prior for the Hubble constant we adopt the standard value $h=0.72 \pm 0.08$ from \citet{Freedman}. Note that some of the priors in equation (\ref{chi_gas_frac}) differ slightly from the ones used in earlier works of \citet{Allen2002,Allen2003,Allen2004} and \citet{Zhu1}. In the following we perform fits with and without the priors in (\ref{chi_gas_frac}).

\begin{figure}
\includegraphics[angle=0,scale=0.4]{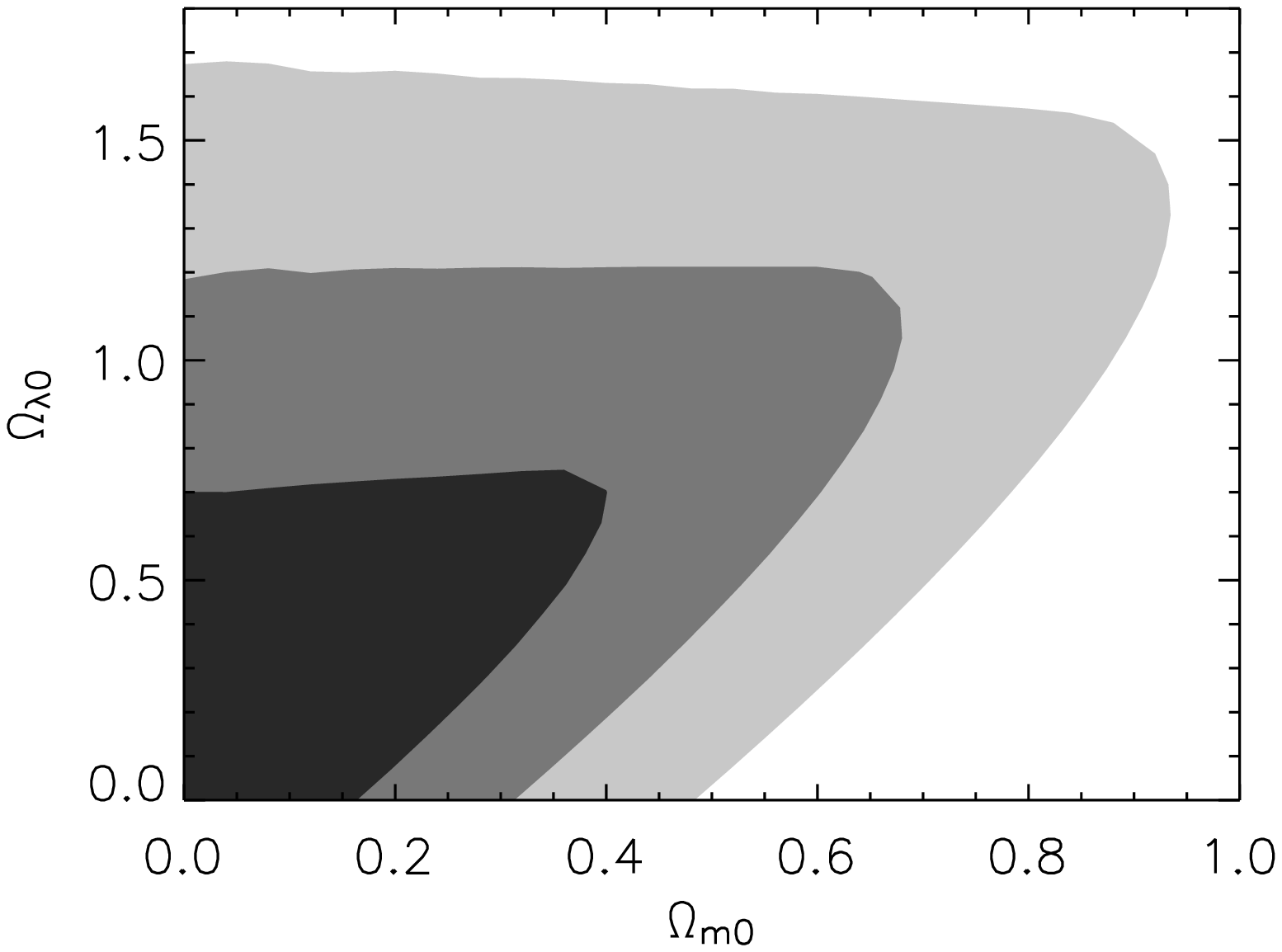}
\includegraphics[angle=0,scale=0.4]{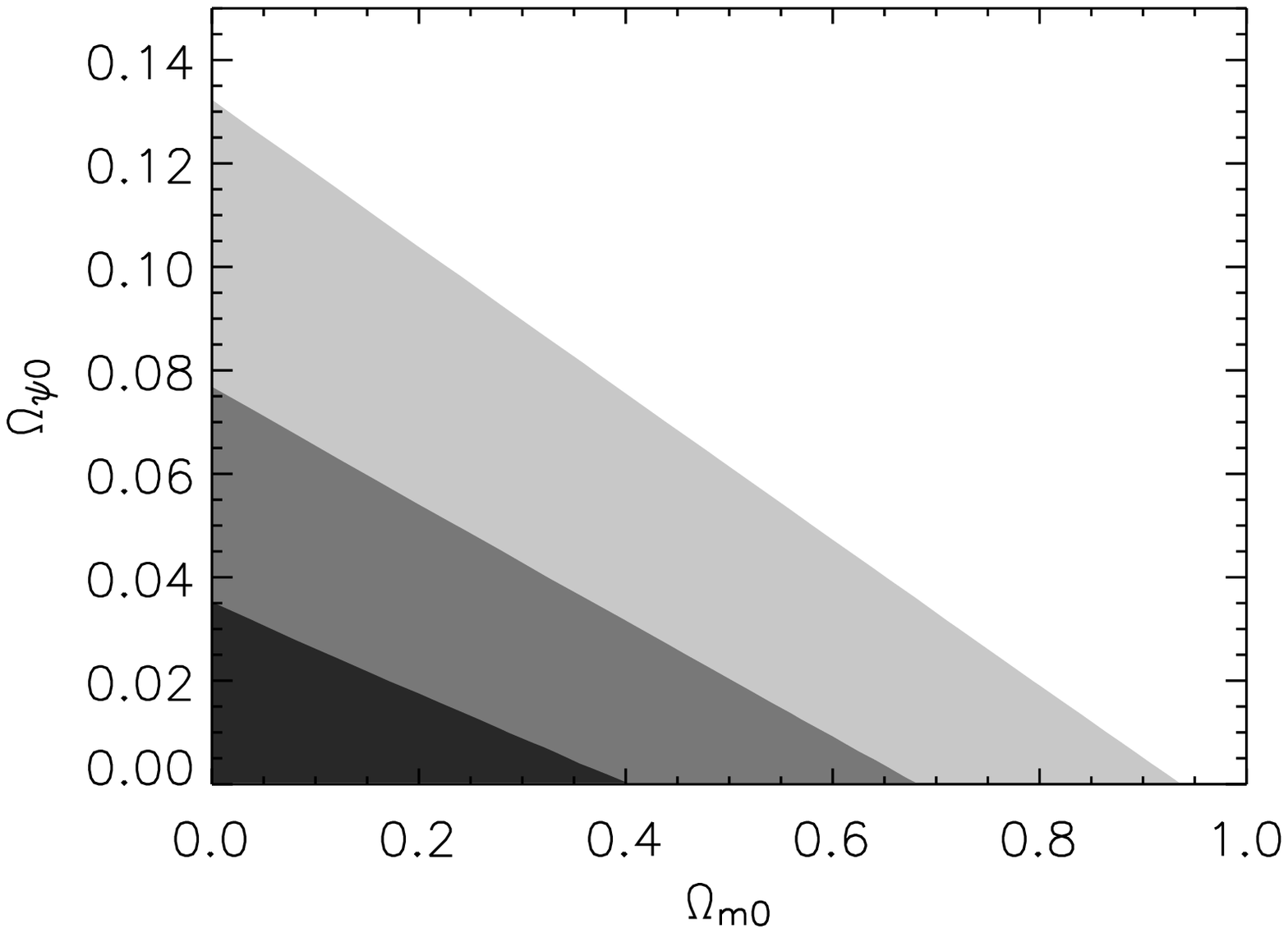}\\
\includegraphics[angle=0,scale=0.4]{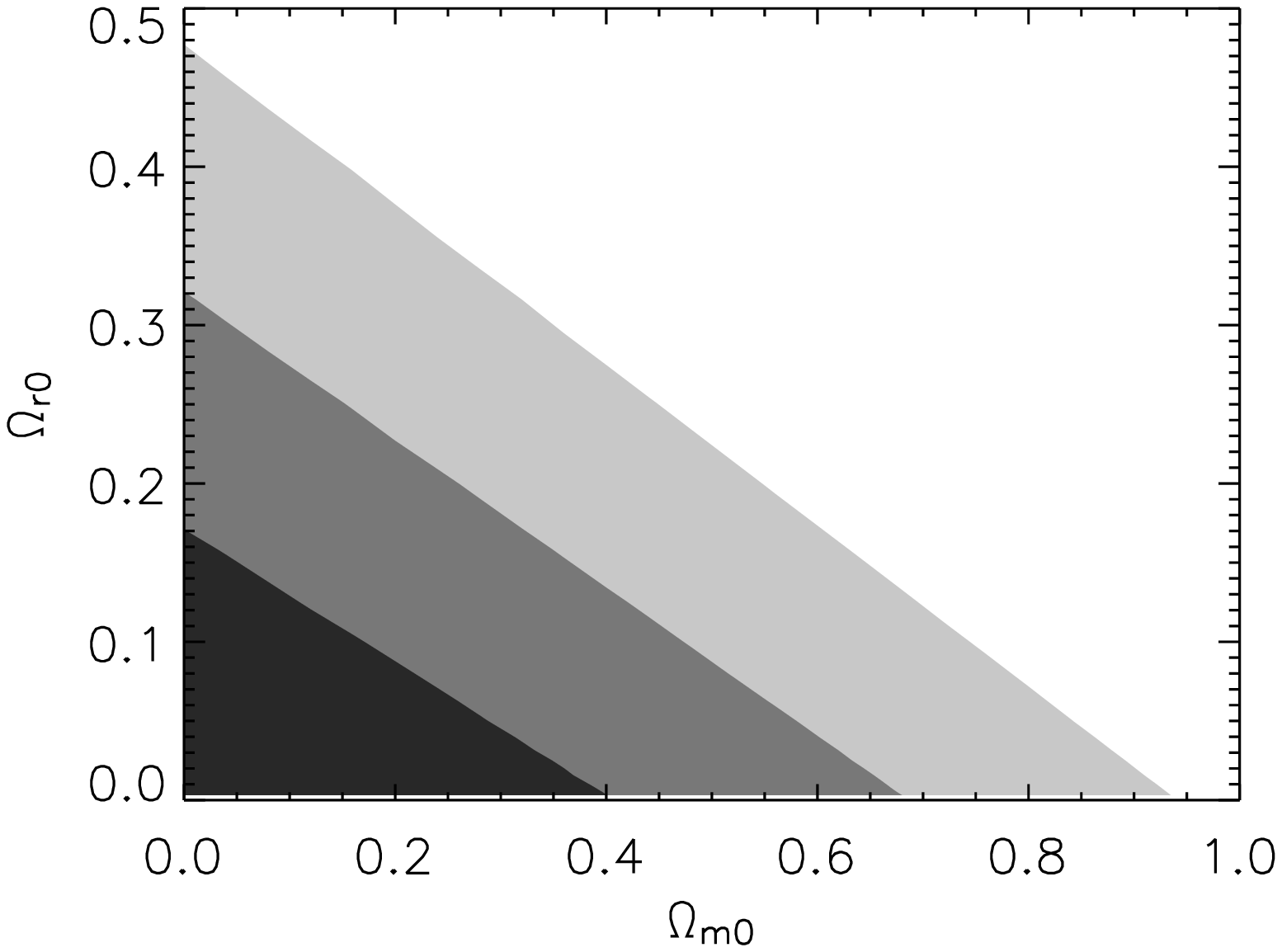}
\includegraphics[angle=0,scale=0.4]{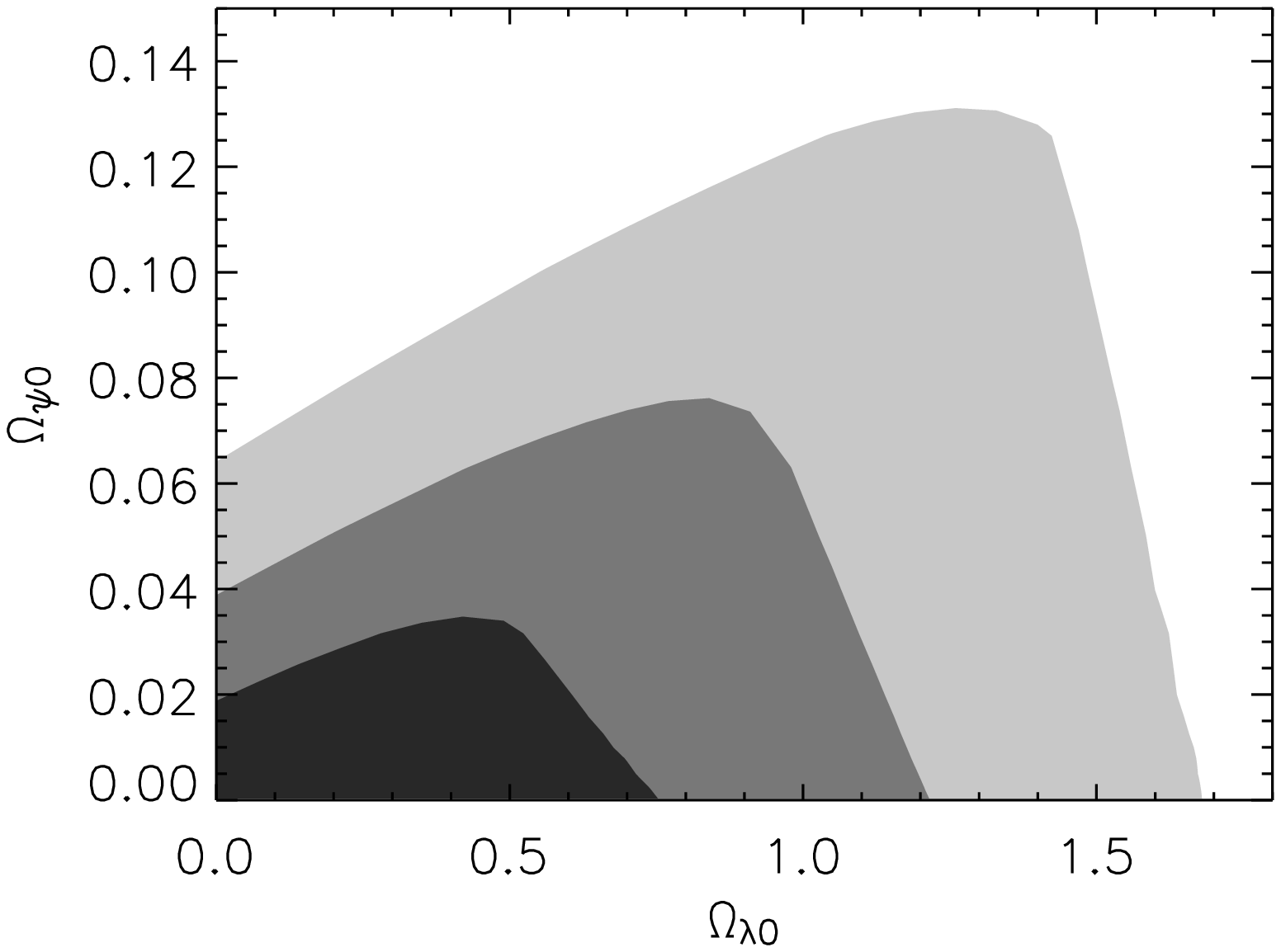}\\
\includegraphics[angle=0,scale=0.4]{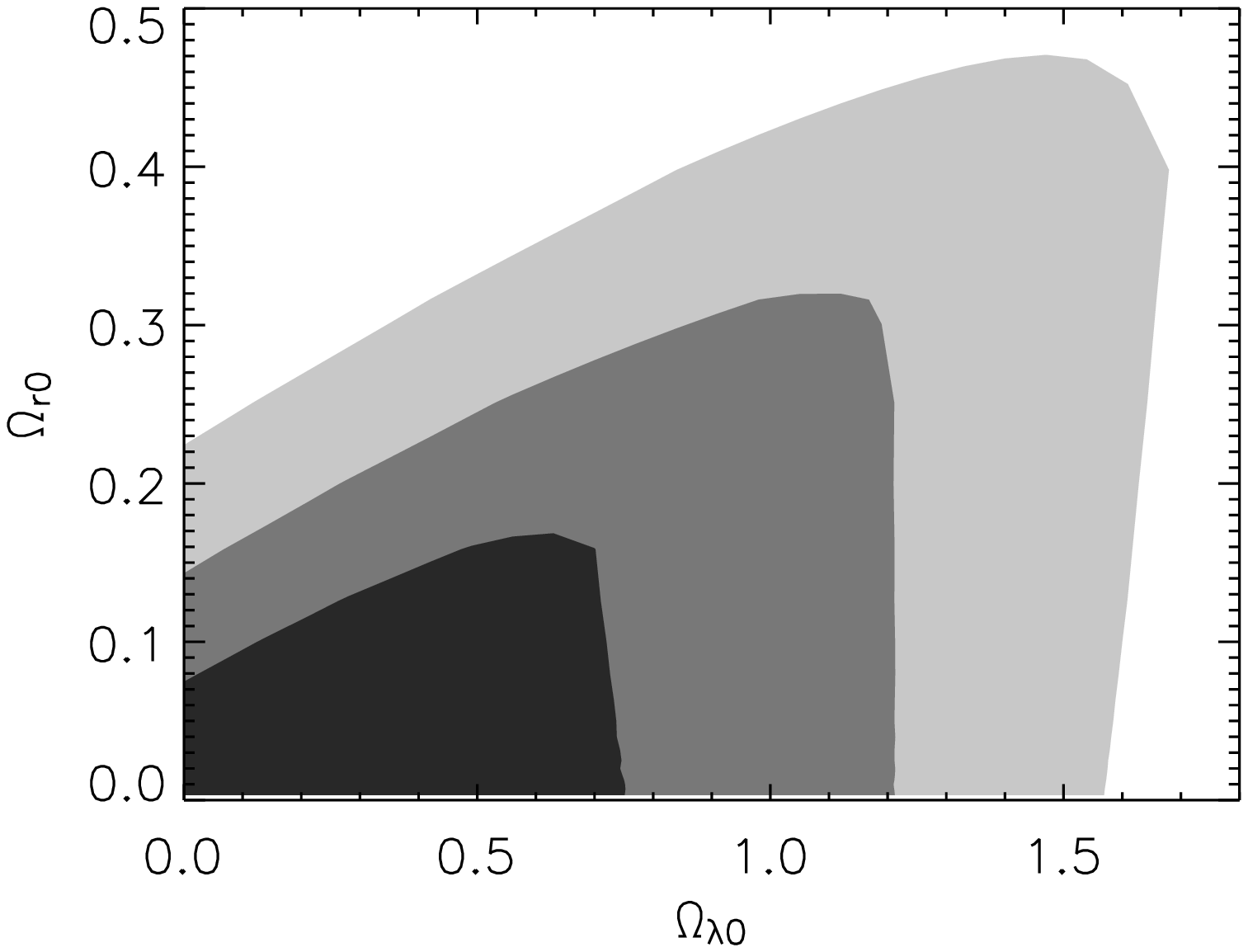}
\includegraphics[angle=0,scale=0.4]{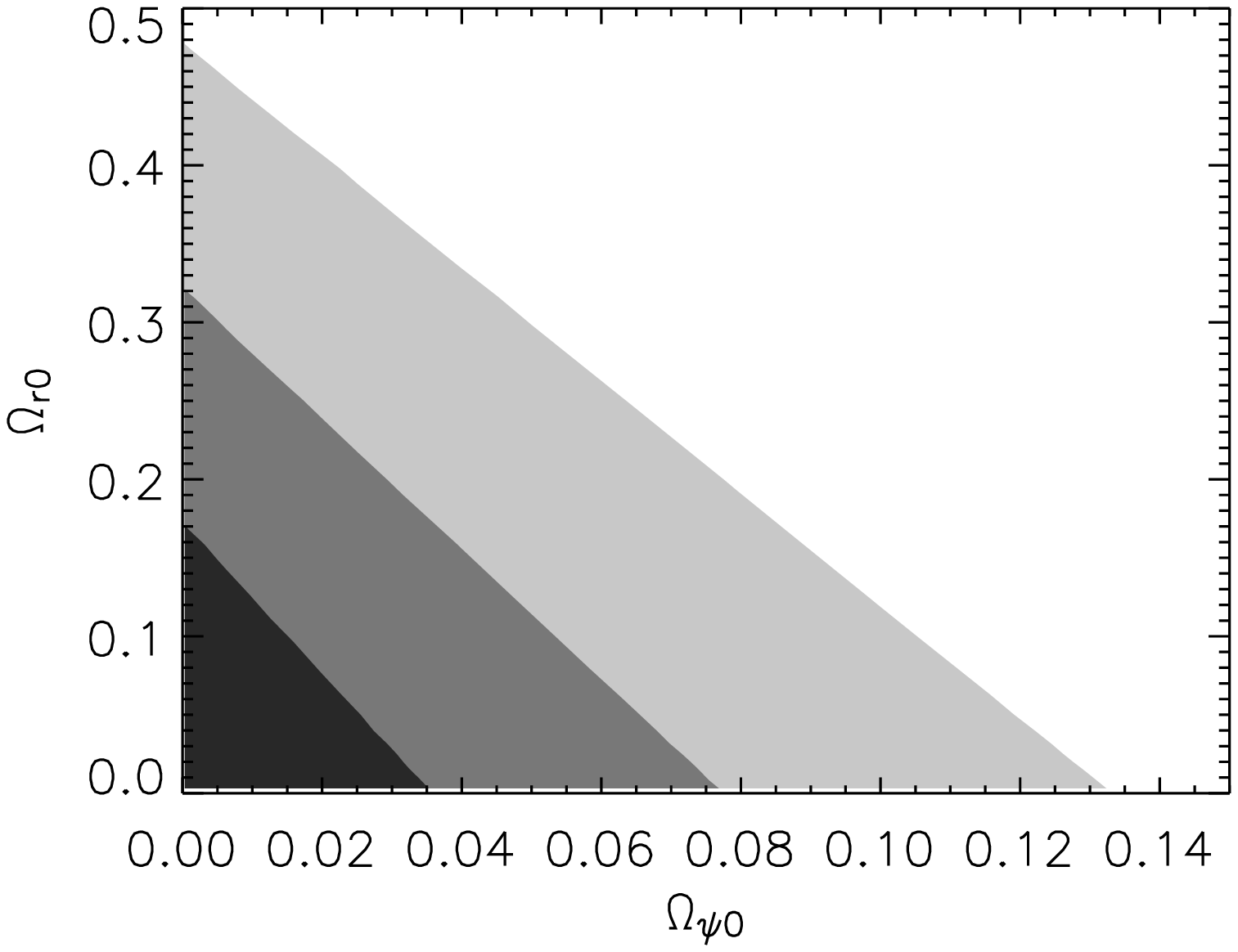}
\caption{Confidence contours (1$\sigma$, 2$\sigma$, 3$\sigma$) for the FR IIb radio galaxy data set of \citet{Daly2003}.}
\label{FIG_FRIIb_confidence_contours}
\end{figure}

\paragraph{Fit results\label{SUB_fit_results}}
Our fit results for the FR IIb radio galaxies are summarized in table \ref{TAB_FRII_fit_results}. There we provide the 1$\sigma$ and 2$\sigma$ confidence limits on the parameters of our model. Additionally, we worked out the 1$\sigma$, 2$\sigma$, and 3$\sigma$ confidence contours in all parameter planes, they are displayed in figure \ref{FIG_FRIIb_confidence_contours}.

\begin{deluxetable}{clll}
\tablecaption{1-d parameter constraints from FR IIb radio galaxies\label{TAB_FRII_fit_results}.}
\tablewidth{0pt}
\tablehead{\colhead{Parameter}&\colhead{Best-fit}&\colhead{$1\sigma$ (68\%)}&\colhead{$2\sigma$ (95\%)}} 
\startdata
$\Omega_{{\rm m}0}$  &$0$&$[0,0.25]$  &$[0,0.54]$\\
$\Omega_{{\lambda}0}$&$0.02$&$[0,0.47]$  &$[0,0.98]$\\
$\Omega_{{\psi}0}$ &$0$&$[0,0.019]$&$[0,0.053]$\\
$\Omega_{{\rm r}0}$&$0$&$[0,0.097]$ &$[0,0.238]$\\
\enddata
\tablecomments{See equation (\ref{chi_square_radio}) for the definition of $\chi^2$, $\chi^2_{\rm min}=18.85$, 16 d.o.f.}
\end{deluxetable}

\placefigure{FIG_FRIIb_confidence_contours}

The 1$\sigma$ and 2$\sigma$ fit results for the clusters are summarized in table \ref{TAB_x_ray_fit_results}. As practiced in case of the FR IIb galaxies we also worked out the 1$\sigma$, 2$\sigma$ and 3$\sigma$ confidence contours in all parameter planes, these are displayed in figure \ref{FIG_Xray_confidence_contours1}. The estimates in table \ref{TAB_x_ray_fit_results} and figure \ref{FIG_Xray_confidence_contours1} correspond to the case {\it without} the priors from equation (\ref{chi_gas_frac}). Since in this case the bias factor $b$ and the Hubble constant $h$ only appear as a prefactor in the expression for the gas mass fraction (\ref{gas_mass_fraction_Allen_et_al}) we combined them in a new prefactor together with the baryonic and pressureless matter mass fractions $A:={\Omega_{b0} \,b\, h^{3/2}} {{\Omega_{{\rm m}0}^{-1}} \left(1+0.19 \sqrt{h}\right)^{-1}}$. With this definition the $\chi^2$ in (\ref{chi_gas_frac}) depends on the following independent parameters $\left(A, \Omega _{{\rm m}0}, \Omega _{\lambda 0},\Omega _{\psi 0},\Omega _{{\rm r}0}\right)$.

\begin{deluxetable}{clll}
\tablecaption{1-d parameter constraints from X-ray cluster data ({\it without} priors)\label{TAB_x_ray_fit_results}.}
\tablewidth{0pt}
\tablehead{\colhead{Parameter}&\colhead{Best-fit}&\colhead{$1\sigma$ (68\%)}&\colhead{$2\sigma$ (95\%)}} 
\startdata
$\Omega_{{\rm m}0}$  &$0$&$[0,0.22]$  &$[0,0.55]$\\
$\Omega_{{\lambda}0}$&$0.63$&$[0.40,0.97]$  &$[0.14,1.43]$\\
$\Omega_{{\psi}0}$ &0$$&$[0,0.022]$&$[0,0.064]$\\
$\Omega_{{\rm r}0}$  &$0$&$[0,0.0768]$ &$[0,0.238]$\\
$A$&$0.0705$&$[0.068,0.073]$&$[0.065,0.077]$\\
\enddata
\tablecomments{See section \ref{SUB_fit_results} for fitting method, $\chi^2_{\rm min}=23.38$, 21 d.o.f, $A:={\Omega_{b0} \,b\, h^{3/2}} {{\Omega_{{\rm m}0}^{-1}} \left(1+0.19 \sqrt{h}\right)^{-1}}$.}
\end{deluxetable}

\placefigure{FIG_Xray_confidence_contours1}
\placefigure{FIG_Xray_confidence_contours2}

\begin{figure}
\includegraphics[angle=0,scale=0.4]{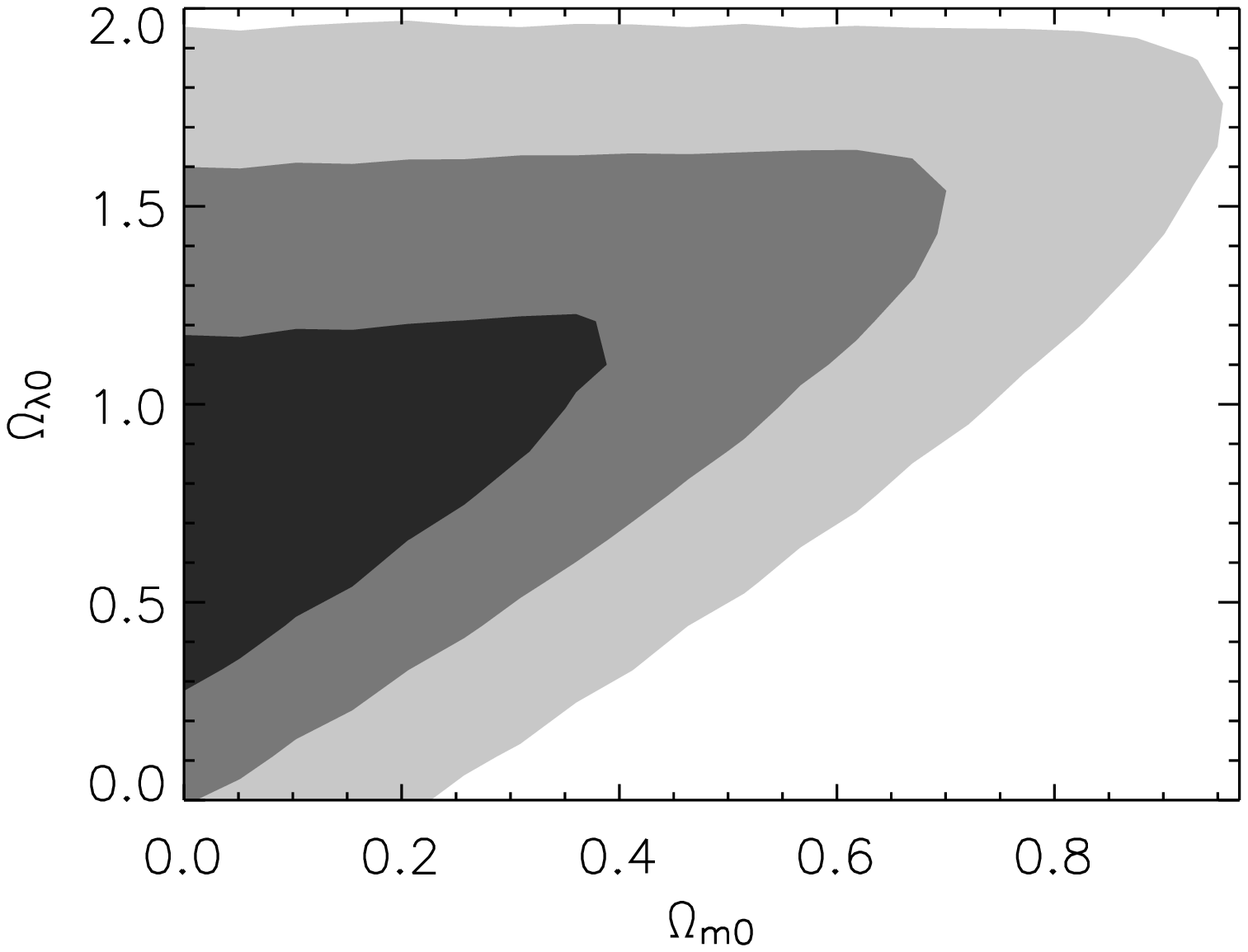}
\includegraphics[angle=0,scale=0.4]{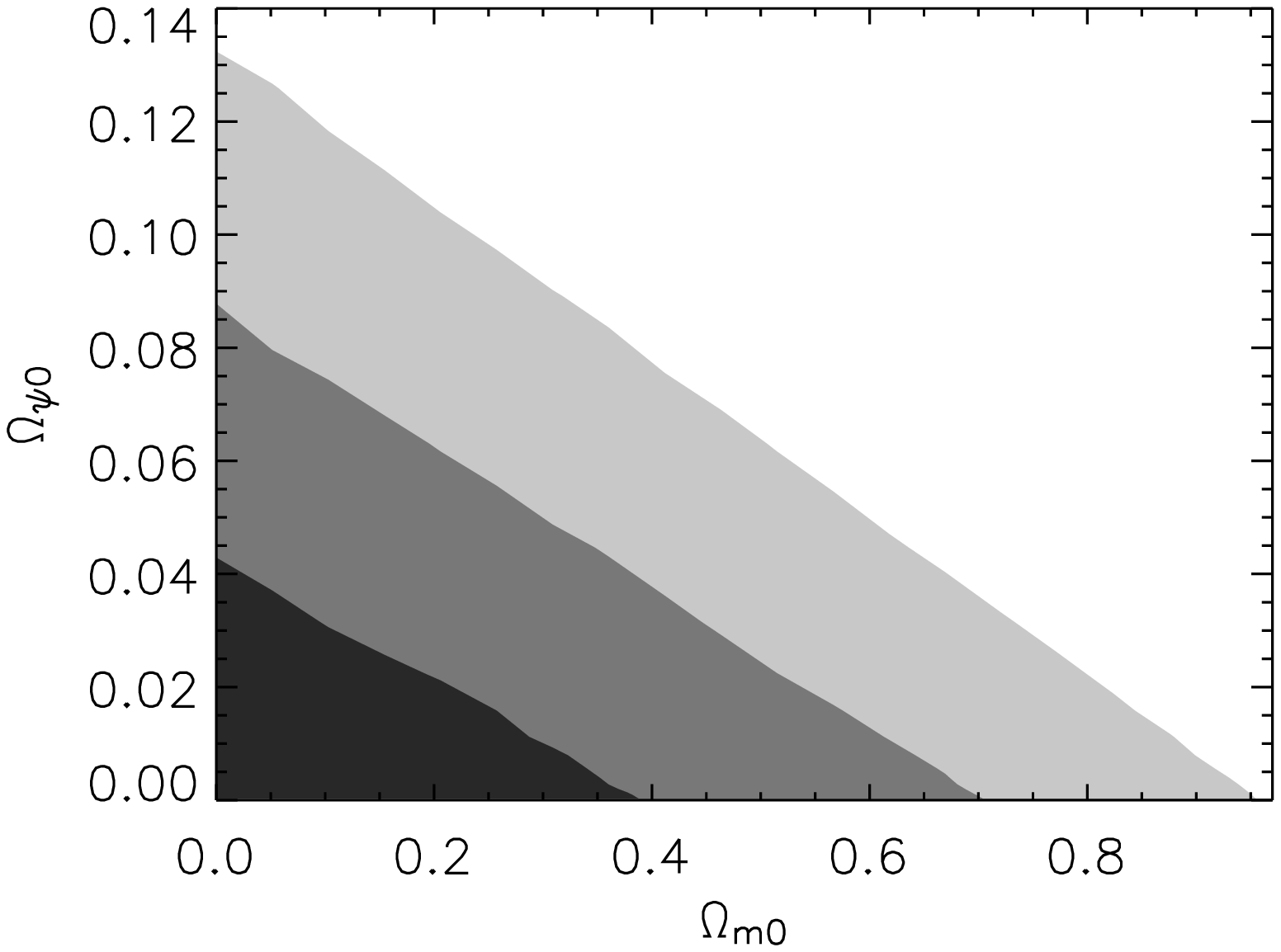}\\
\includegraphics[angle=0,scale=0.4]{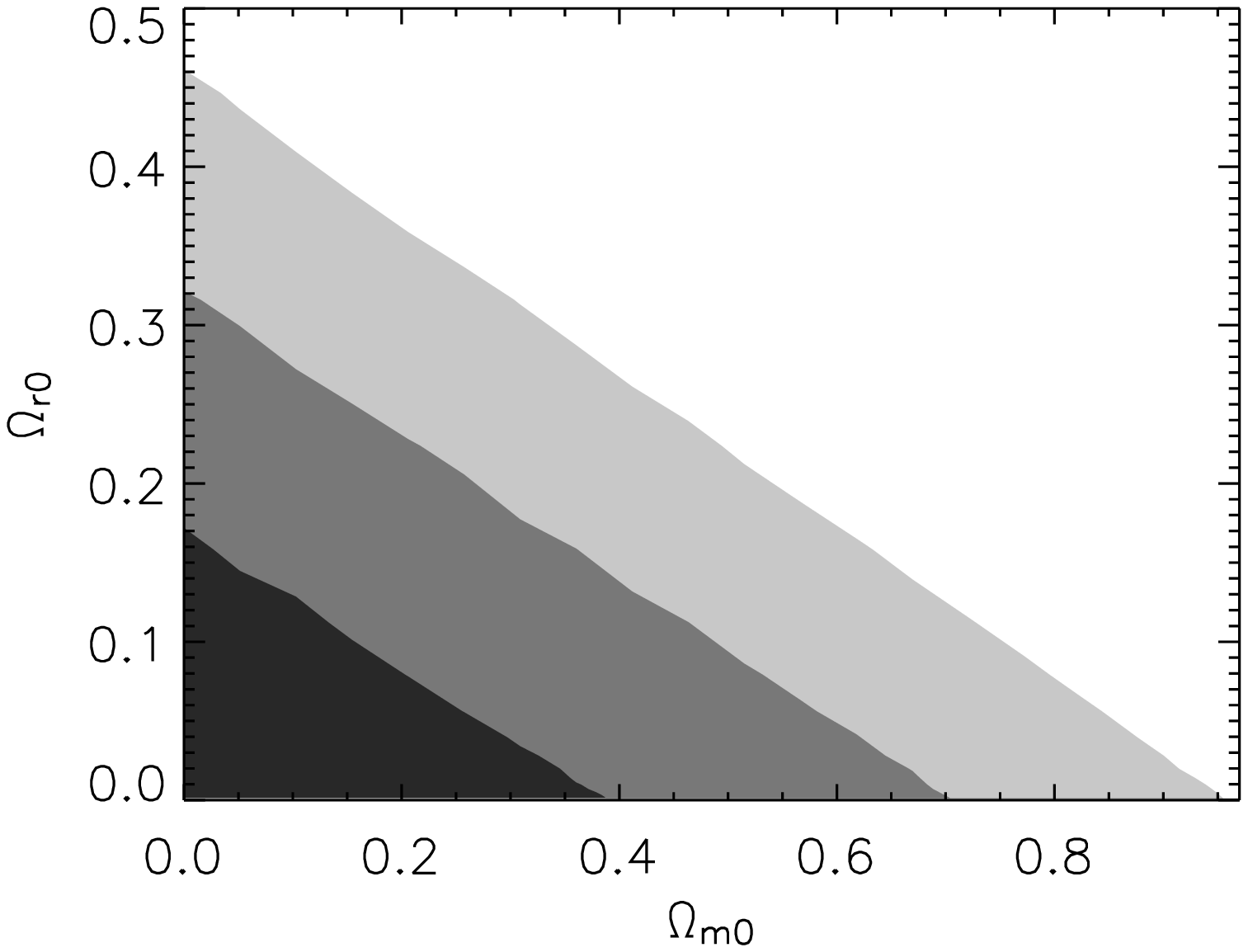}
\includegraphics[angle=0,scale=0.4]{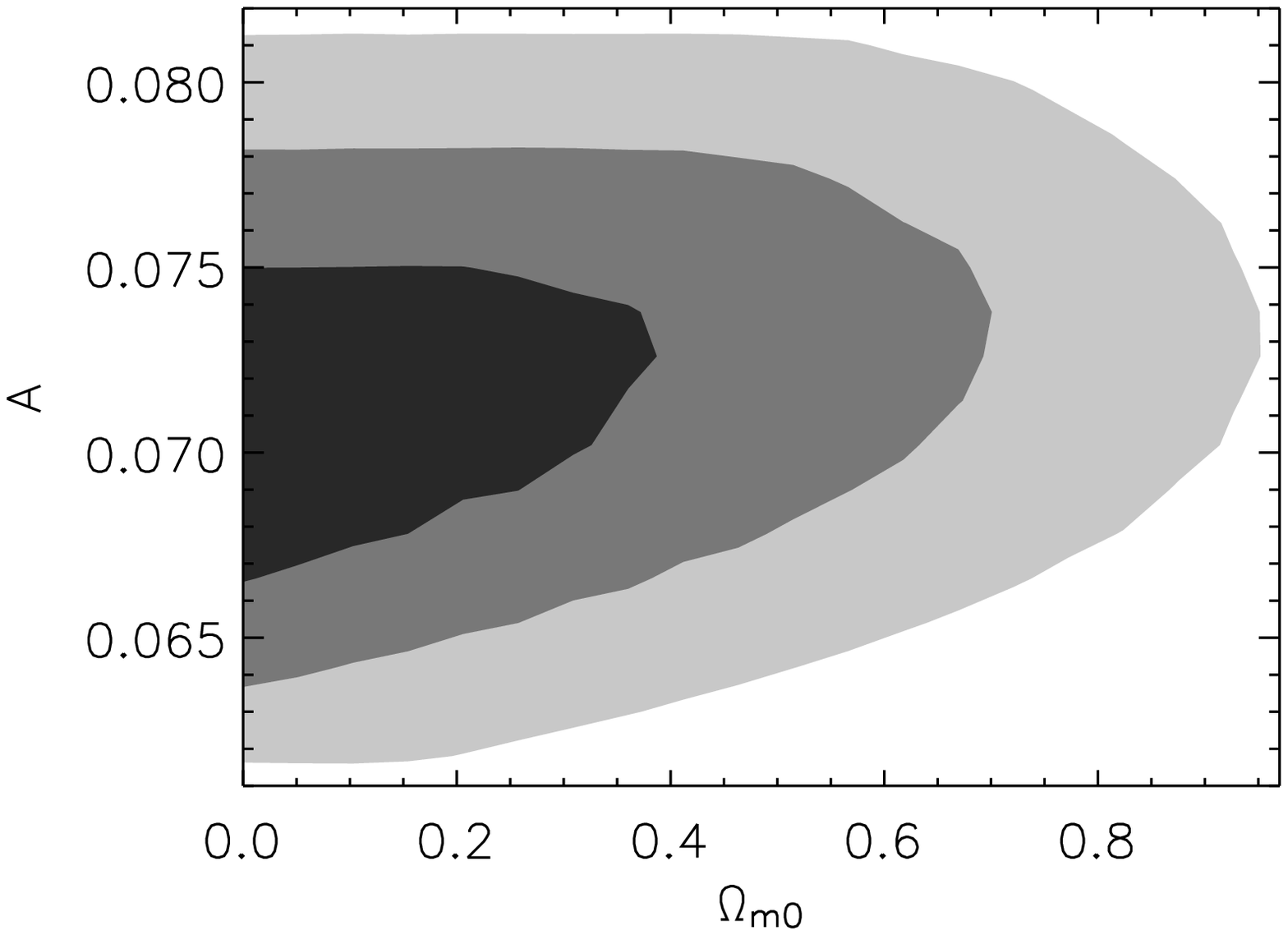}\\
\includegraphics[angle=0,scale=0.4]{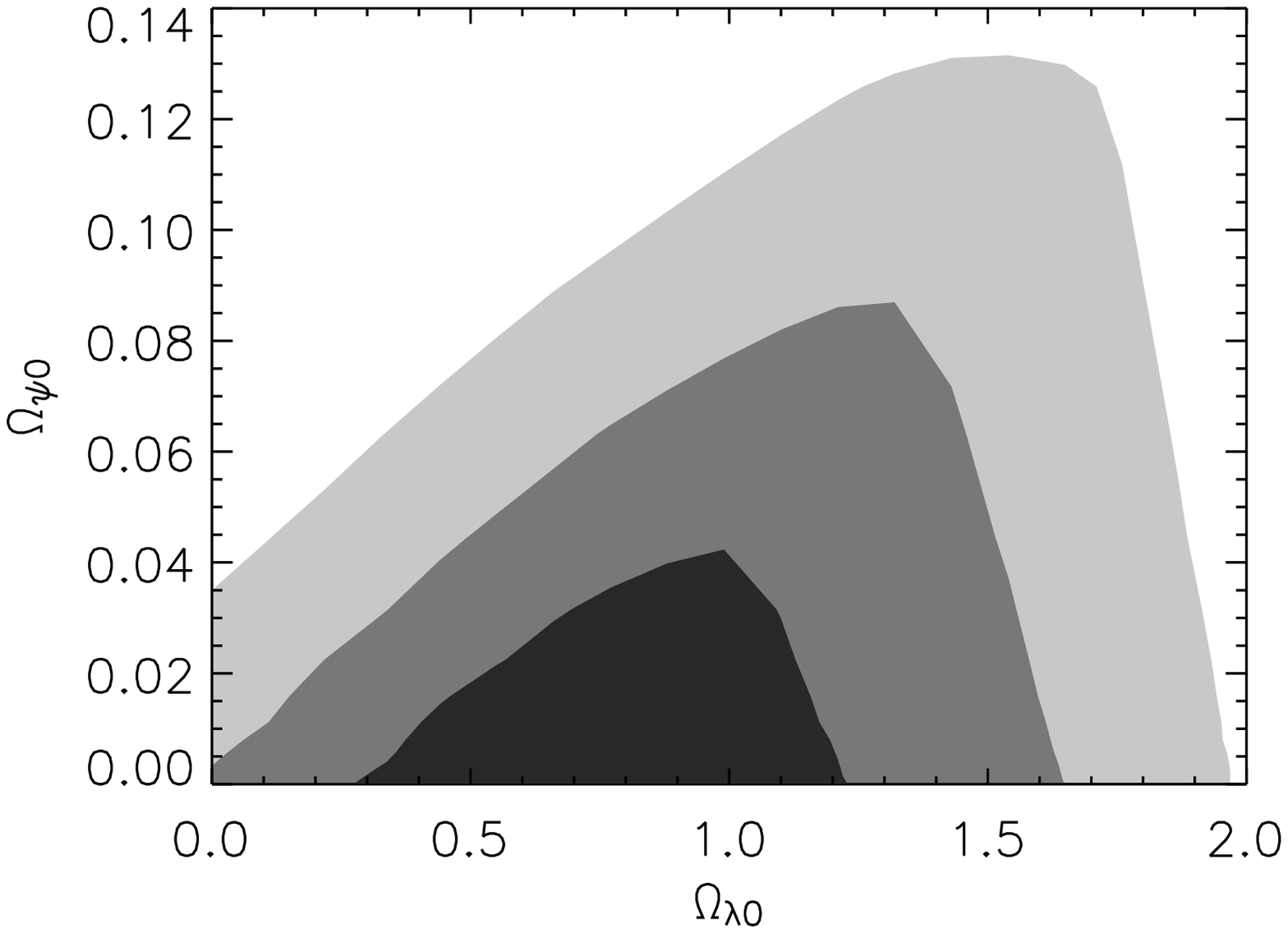}
\includegraphics[angle=0,scale=0.4]{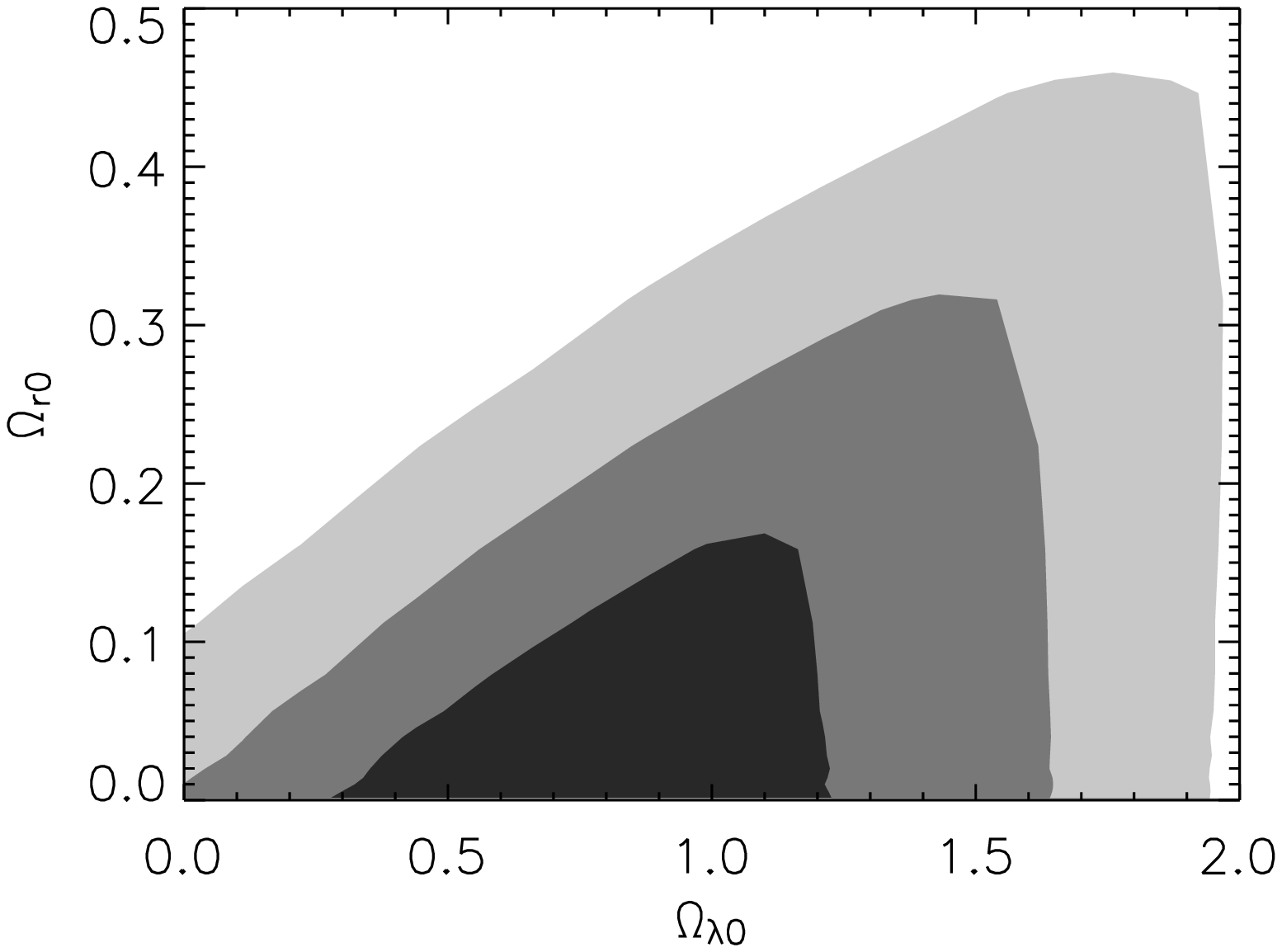}
\caption{Confidence contours (1$\sigma$, 2$\sigma$, 3$\sigma$) for the gas mass fraction data of \citet{Allen2004} ({\it without} priors). $A:={\Omega_{b0} \,b\, h^{3/2}} {{\Omega_{{\rm m}0}^{-1}} \left(1+0.19 \sqrt{h}\right)^{-1}}$.}
\label{FIG_Xray_confidence_contours1}
\end{figure}

\begin{figure}
\addtocounter{figure}{-1}
\includegraphics[angle=0,scale=0.4]{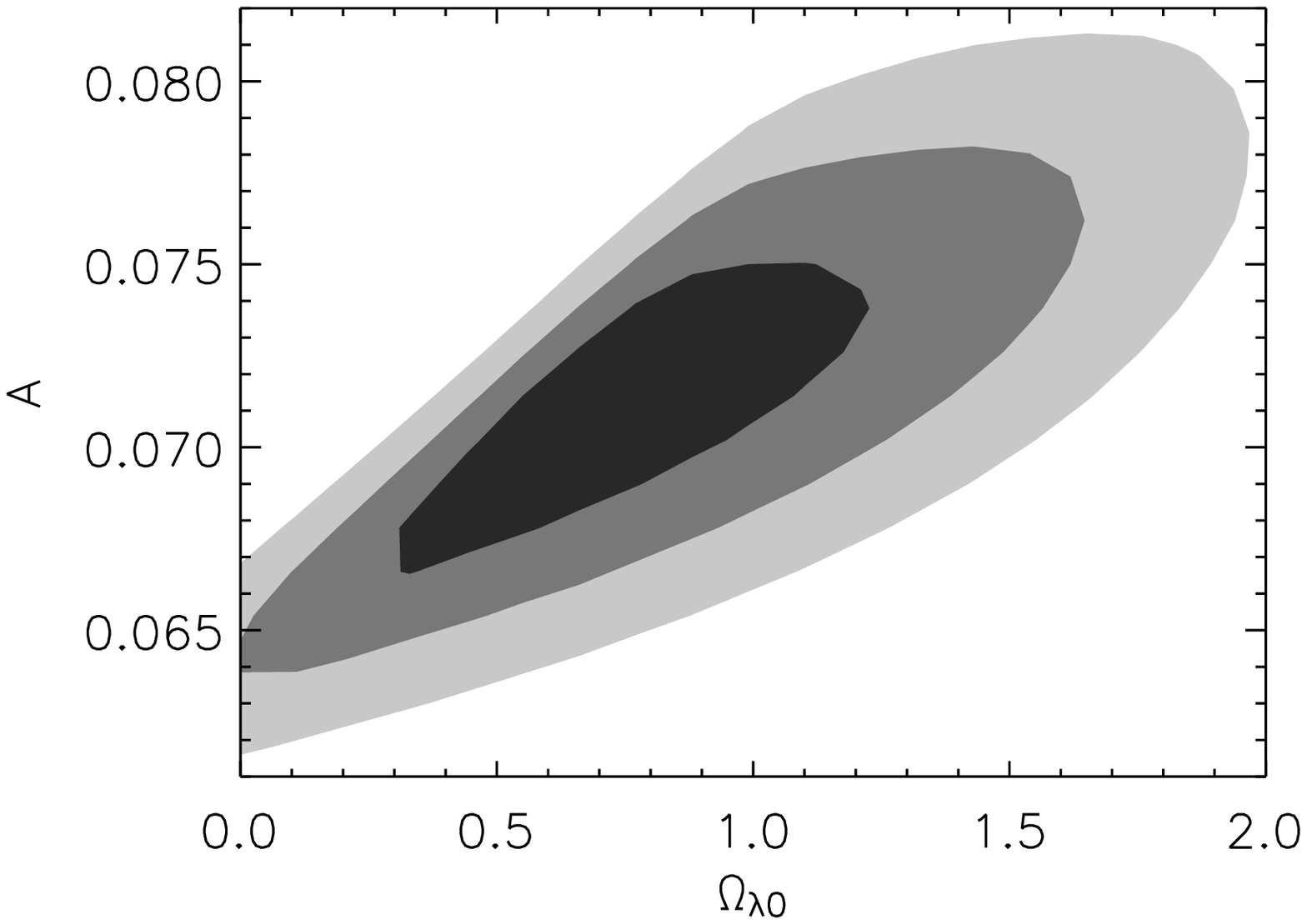}
\includegraphics[angle=0,scale=0.4]{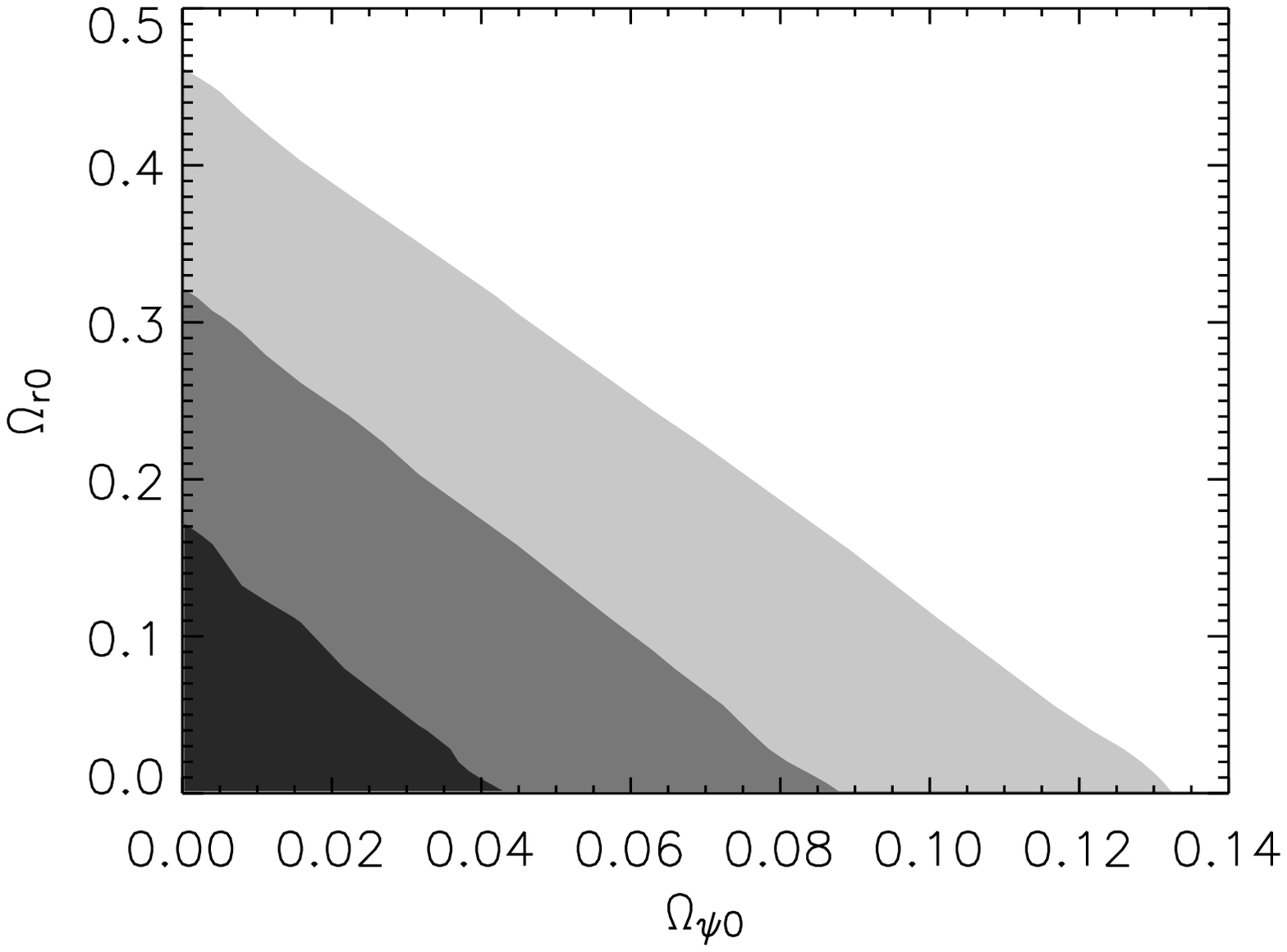}\\
\includegraphics[angle=0,scale=0.4]{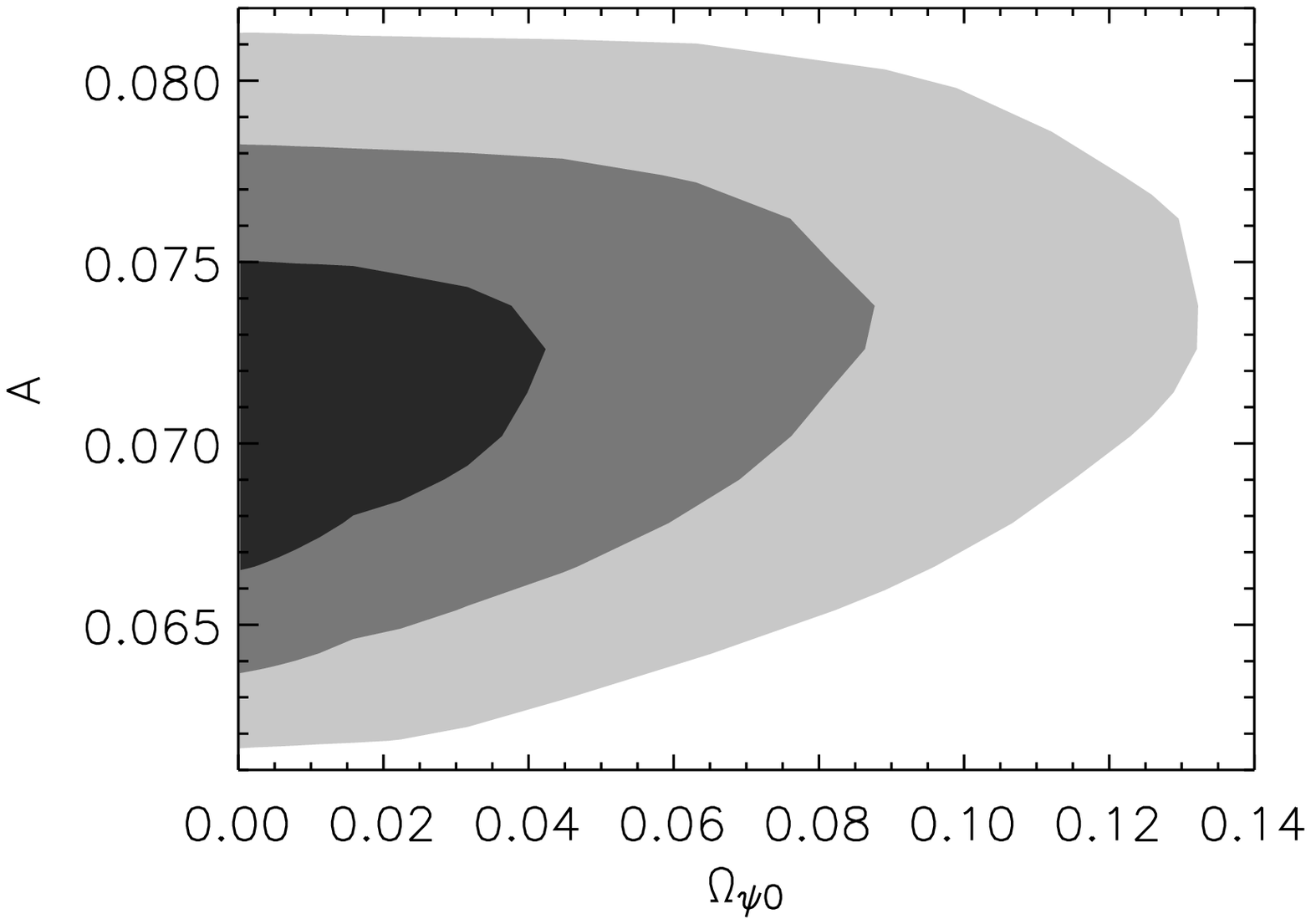}
\includegraphics[angle=0,scale=0.4]{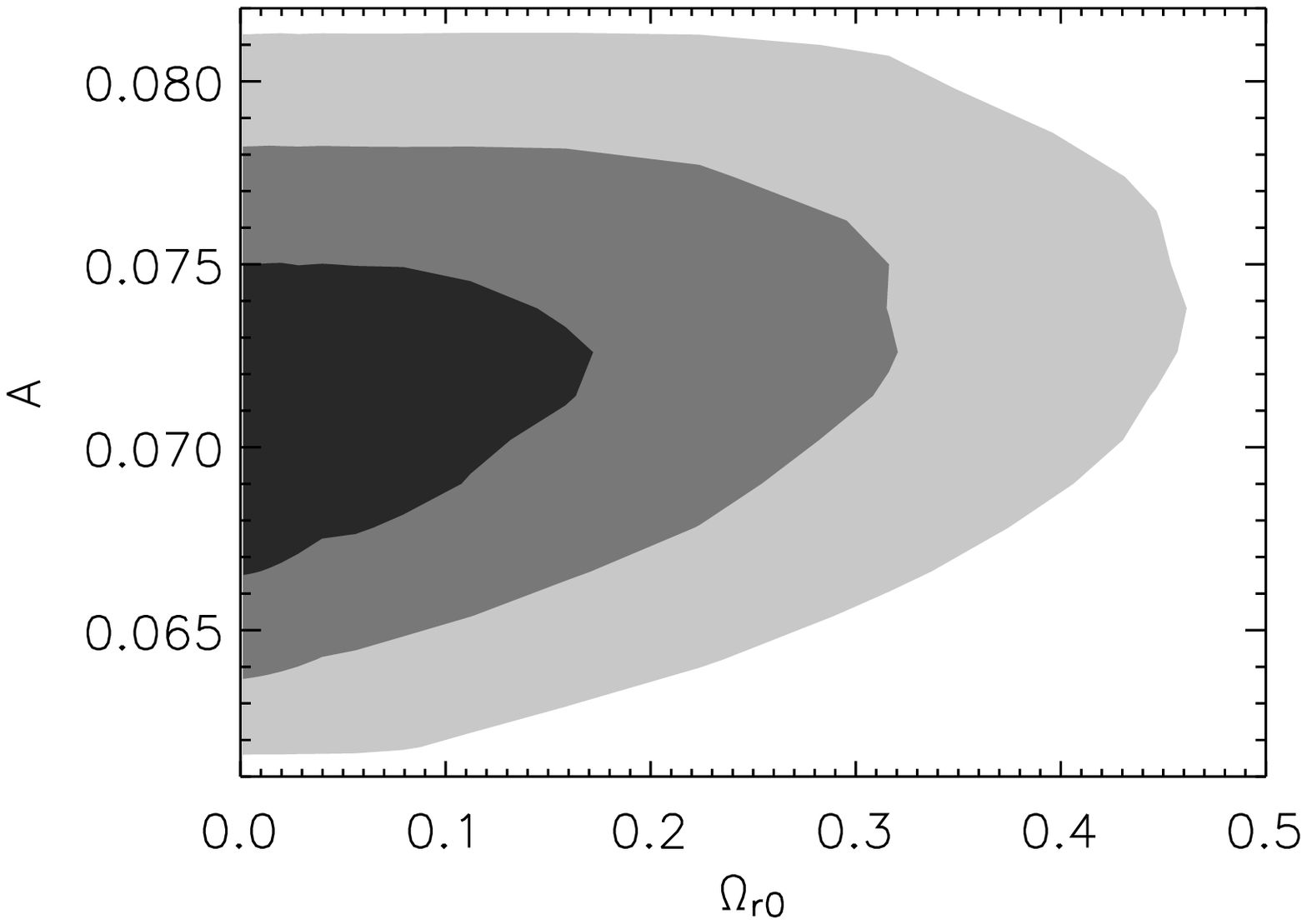}
\caption{(continued)}
\label{FIG_Xray_confidence_contours2}
\end{figure}

In table \ref{TAB_combined_results} we list the 1$\sigma$ and 2$\sigma$ confidence intervals for a simultaneous fit to the FR IIb radio galaxies and X-ray gas mass fractions in clusters. Again we do {\it not} use any priors in our fitting procedure. 

\begin{deluxetable}{clll}
\tablecaption{1-d parameter constraints for combined FR IIb/X-ray data set ({\it without} priors)\label{TAB_combined_results}.}
\tablewidth{0pt}
\tablehead{\colhead{Parameter}&\colhead{Best-fit}&\colhead{$1\sigma$ (68\%)}&\colhead{$2\sigma$ (95\%)}} 
\startdata
$\Omega_{{\rm m}0}$ &$0$ &$[0,0.22]$  &$[0,0.44]$\\
$\Omega_{{\lambda}0}$&$0.25$&$[0.07,0.61]$  &$[0,0.99]$\\
$\Omega_{{\psi}0}$&$0.003$ &$[0,0.021]$&$[0,0.045]$\\
$\Omega_{{\rm r}0}$&$0$  &$[0,0.09]$ &$[0,0.19]$\\
$A$&$0.066$&$[0.064,0.069]$&$[0.062,0.071]$\\
\enddata
\tablecomments{See section \ref{SUB_fit_results} for fitting method, $\chi^2_{\rm min}=46.06$, 41 d.o.f, $A:={\Omega_{b0} \,b\, h^{3/2}} {{\Omega_{{\rm m}0}^{-1}} \left(1+0.19 \sqrt{h}\right)^{-1}}$.}
\end{deluxetable}

Finally, in table \ref{TAB_x_ray_with_assumptions_results} we provide the 1$\sigma$ and 2$\sigma$ confidence intervals for a fit to the X-ray gas mass fractions with the priors from equation (\ref{chi_gas_frac}). 

All of the fits without priors prefer a universe with a very low pressureless matter component (in contrast to what is assumed within the usual cosmological concordance model with $\Omega_{{\rm m}0} \approx 0.3$). At the 2$\sigma$ level the estimates for $\Omega_{{\rm m}0}$ from FR IIb galaxies and the clusters nearly equal each other. The cluster data seem to prefer slightly higher values of the cosmological constant. A quick glance at figures \ref{FIG_FRIIb_confidence_contours} and \ref{FIG_Xray_confidence_contours1} reveals that there is a sufficient overlap of the confidence contours in the different parameter planes. This result is reassuring since the two different methods considered here yield compatible estimates for the cosmological parameters.

The introduction of priors from HST, BBN, and cluster simulations leads to a shift of the preferred parameters to higher values of $\Omega_{{\rm m}0}$ and $\Omega_{{\lambda}0}$. The confidence limits on these two parameters are in complete agreement with the results of \citet{Allen2004}, if one takes into account that we perform fits with two additional cosmological parameters. This brings us to the main concern of this work, namely to constrain the new density parameter $\Omega_{\psi 0}$ with the help of the FR IIb and X-ray cluster data. From all of our fits we can infer an upper limit on $\Omega_{\psi 0}$ which cannot account for more than 10\% of the critical density. This is a rather generous upper limit, in fact one expects that the present-day value of the new component is much closer to zero than inferred from the FR IIb and cluster data alone, the main reason for that is the $\sim z^6$ scaling behaviour which leads to a too early domination at higher redshifts which are currently probed by the CMB and nucleosynthesis.
 
\begin{deluxetable}{clll}
\tablecaption{1-d parameter constraints from X-ray clusters with priors for $b, h$, and $\Omega_{{\rm b} 0}$\label{TAB_x_ray_with_assumptions_results}.}
\tablewidth{0pt}
\tablehead{\colhead{Parameter}&\colhead{Best-fit}&\colhead{$1\sigma$ (68\%)}&\colhead{$2\sigma$ (95\%)}} 
\startdata
$\Omega_{{\rm m}0}$ &$0.25$ &$[0.21,0.28]$  &$[0.18,0.33]$\\
$\Omega_{{\lambda}0}$&$0.97$&$[0.78,1.16]$  &$[0.50,1.48]$\\
$\Omega_{{\psi}0}$&$0$ &$[0,0.012]$&$[0,0.040]$\\
$\Omega_{{\rm r}0}$&$0$  &$[0,0.047]$ &$[0,0.15]$\\
$\Omega_{{\rm b}0}$&$0.040$  &$[0.034,0.050]$ &$[0.027,0.065]$\\
$b$&$0.82$&$[0.74,0.89]$&$[0.65,0.98]$\\
$h$&$0.73$&$[0.66,0.80]$&$[0.60,0.87]$\\
\enddata
\tablecomments{See section \ref{SUB_fit_results} for fitting method, $\chi^2_{\rm min}=24.50$, 22 d.o.f.}
\end{deluxetable}

\section{Conclusions}\label{CONCLUSION_section}

In this work we used recent data on FR IIb radio galaxies and X-ray gas mass fractions of relaxed clusters to constrain the parameters within an alternative cosmological model. We were able to place an upper limit on the new density parameter $\Omega_{\psi 0}$ which controls the non-Riemannian features of the underlying cosmological model. 

\paragraph{Deceleration parameter \& age}

\begin{figure}
\includegraphics[angle=-90,scale=0.55]{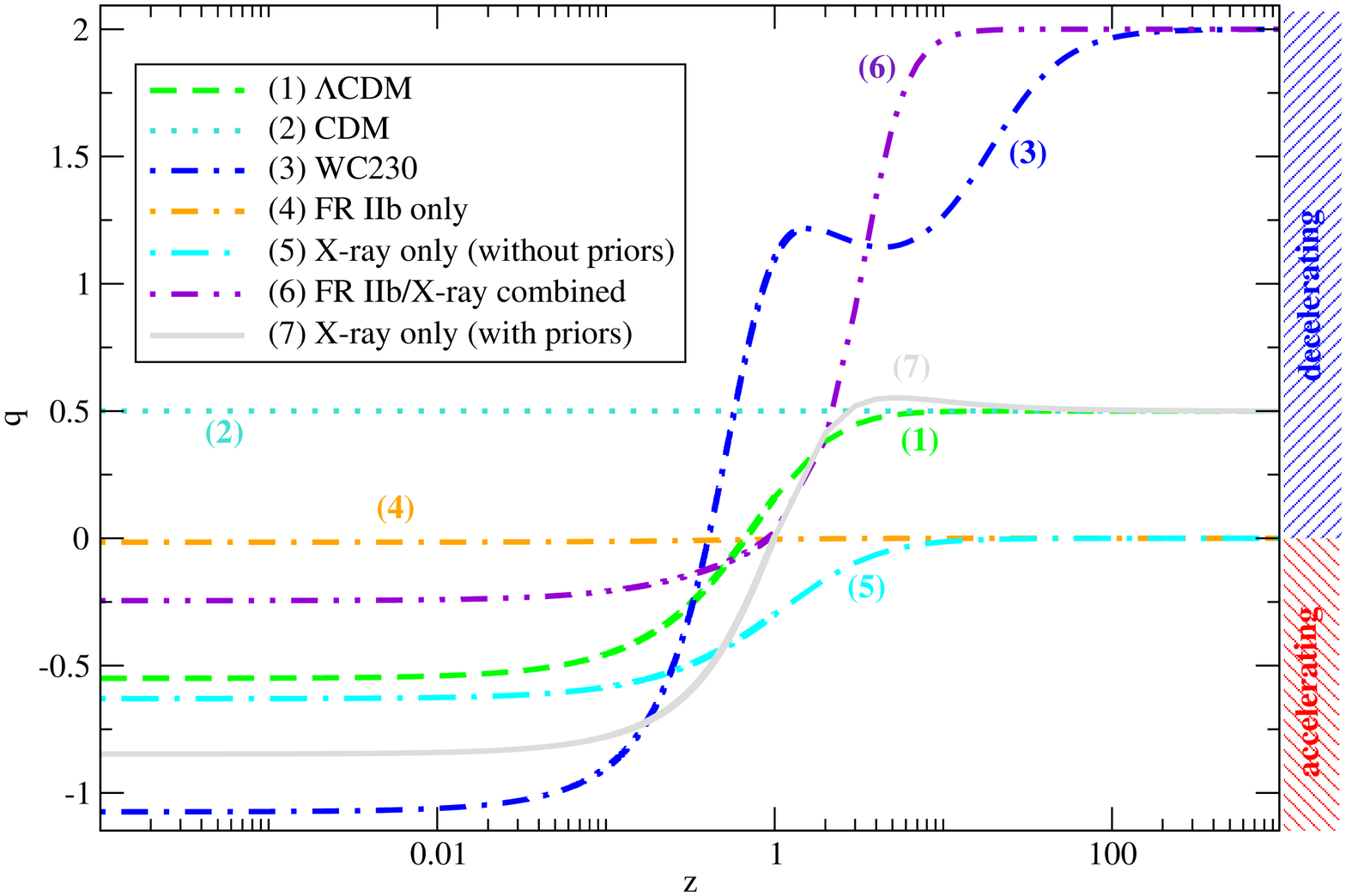}
\caption{Deceleration for various models, cf.\ tables \ref{TAB_FRII_fit_results}--\ref{TAB_x_ray_with_assumptions_results}. The nomenclature WC230 corresponds to the one used in table 2 of \citet{PuetzfeldChen}. By CDM we denote the model with $\Omega_{{\rm m}0}=1$, and $\Lambda$CDM corresponds to the popular choice $\Omega_{{\rm m}0}=0.3$, $\Omega_{{\lambda}0}=0.7$, all other parameters are set to zero.}\label{FIG_deceleration_factor}
\end{figure}

In figure \ref{FIG_deceleration_factor} we plotted the deceleration parameter $q$ versus the redshift for the best-fit parameters provided in tables \ref{TAB_FRII_fit_results}-\ref{TAB_x_ray_with_assumptions_results}, cf.\ also equation (17) in \citet{PuetzfeldChen}. As becomes clear from the plot all of the models, except the old CDM model, support the notion that the universe is presently undergoing an accelerated phase of expansion. 

\placefigure{FIG_decleration_factor}

\begin{figure}
\includegraphics[angle=-90,scale=0.55]{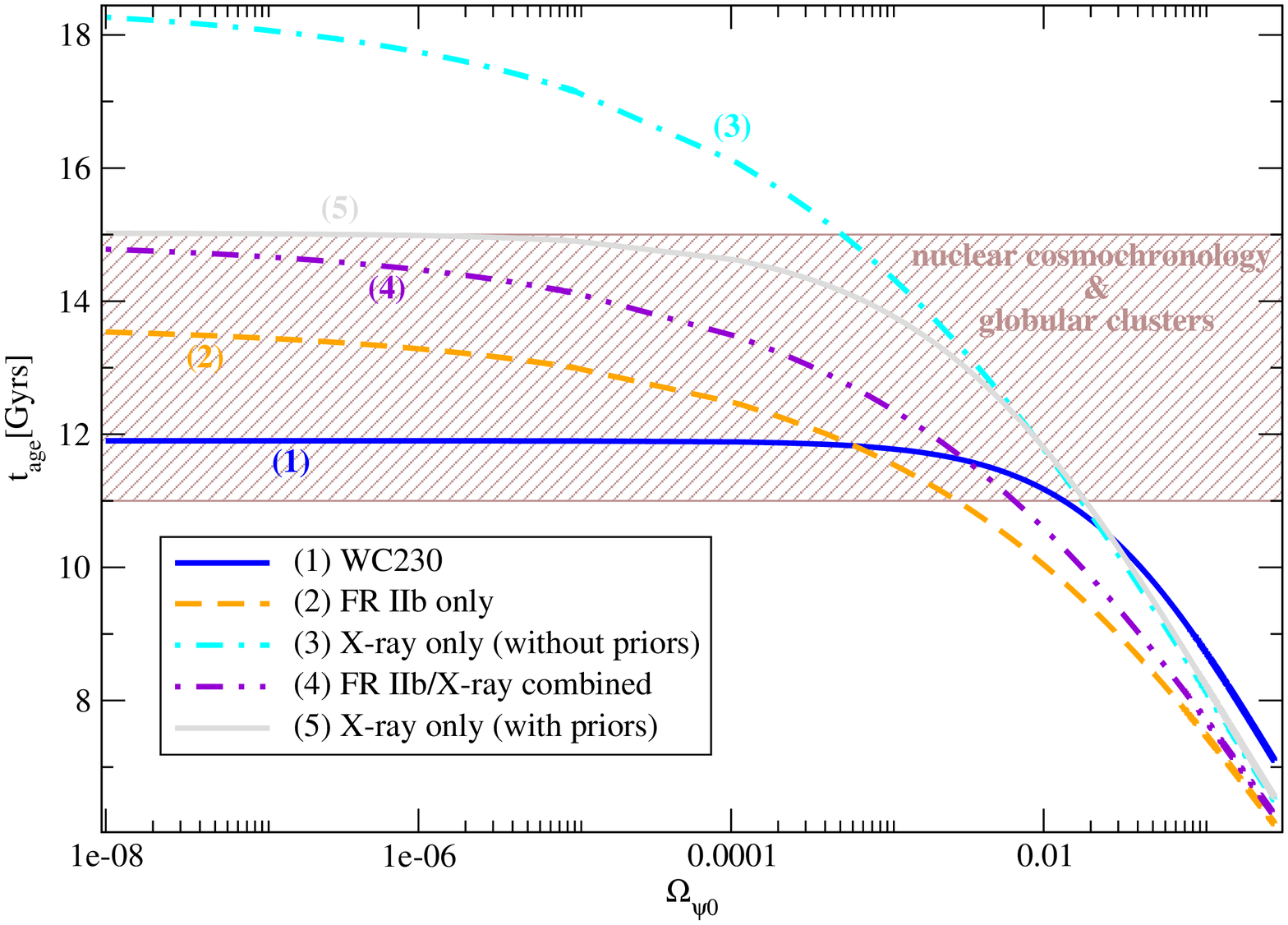}
\caption{Age for various models, cf.\ tables \ref{TAB_FRII_fit_results}--\ref{TAB_x_ray_with_assumptions_results}. The nomenclature WC230 corresponds to the one used in table 2 of \citet{PuetzfeldChen}. For the best-fit sets from tables \ref{TAB_FRII_fit_results}-\ref{TAB_combined_results} we adopted the standard value $h=0.72$.}
\label{FIG_age}
\end{figure}

In figure \ref{FIG_age} we display the variation of the age of the universe for different values of the new density parameter $\Omega_{\psi 0}$, the rest of the parameters being fixed to their best-fit values, in comparison to the age estimates from globular clusters and nuclear cosmochronology, which range from 11 to 15 Gyrs \citep{Chaboyer,Truran}. The comparison with the age estimates leads to a more stringent upper limit on $\Omega_{\psi 0}$, which then should not account for more than $1\%$ of the critical density. 

\placefigure{FIG_age}

\paragraph{Comparison with the SNe Ia}
The parameter estimates within this work are compatible with the ones from supernovae of type Ia. The size and shape of the confidence regions in figure \ref{FIG_FRIIb_confidence_contours} and figure \ref{FIG_Xray_confidence_contours1} resemble the ones given in \citet{PuetzfeldChen} for the SN Ia data. The convergence of the parameter estimates from various astrophysical sources, i.e.\ supernovae, radio galaxies, and clusters, is very encouraging especially if one considers the different redshift ranges covered by these data sets.

\paragraph{Outlook}
In future work it would be interesting to make use of the recently released extended SN Ia data set of \citet{Riess2004}. This new data set contains now several SN Ia with redshifts $z>1.25$, thereby covering a range which was formerly unexplored by SNe Ia, if one disregards SN1997ff, thereby allowing a direct comparison with the FR IIb results in that redshift range. 

\bigskip
 
\acknowledgments
The authors are indebted to S.W.\ Allen et al.\ for providing their gas mass fractions data set. Several suggestions of M.K.\ Daniel concerning the readability of this manuscript are gratefully acknowledged. 

\appendix

\section{Data sets}\label{APP_data_sets}

\begin{deluxetable}{ccc}
\tablecaption{FR IIb radio galaxy data. \label{TAB_radio_galaxy_data}}
\tablewidth{0pt}
\tablehead{\colhead{$z$}&\colhead{Coordinate-distance}&\colhead{Error}} 
\startdata
0.056&    0.056&              0.010\\  
0.430&    0.445&              0.071\\  
0.549&    0.400&              0.066\\  
0.572&    0.319&              0.051\\  
0.630&    0.600&              0.071\\  
0.720&    0.606&              0.071\\  
0.749&    0.625&              0.069\\  
0.811&    0.667&              0.081\\  
0.860&    0.818&              0.149\\  
0.967&    0.681&              0.108\\  
0.974&    0.780&              0.127\\  
0.996&    0.703&              0.111\\  
1.079&    0.842&              0.151\\  
1.144&    0.753&              0.126\\  
1.190&    1.141&              0.205\\  
1.210&    0.996&              0.251\\  
1.480&    0.849&              0.206\\  
1.575&    1.477&              0.386\\  
1.681&    1.167&              0.249\\  
1.790&    1.246&              0.257\\  
\enddata
\tablecomments{Data taken from \cite{Daly2003}.}
\end{deluxetable}

\begin{deluxetable}{cccl}
\tablecaption{X-ray cluster data. \label{TAB_x_ray_data_new}}
\tablewidth{0pt}
\tablehead{\colhead{$z$}&\colhead{$f_{\rm gas}$}&\colhead{Error}&\colhead{Source}} 
\startdata
0.077900& 0.188964& 0.011336&    Abell 2029\\          
0.088200& 0.183744& 0.011119&    Abell 478\\        
0.142700& 0.167000& 0.018561&    Abell 1413\\          
0.188000& 0.169000& 0.011180&    Abell 383\\           
0.206000& 0.180000& 0.014509&    Abell 963\\           
0.208000& 0.137000& 0.017720&    RXJ0439.0+0520\\      
0.240000& 0.163000& 0.008631&    4C55\\                
0.252300& 0.164000& 0.011511&    Abell 1835\\          
0.288000& 0.149000& 0.017117&    Abell 611\\           
0.313000& 0.169000& 0.009513&    MS2137.3-2353\\       
\tablenotemark{a}& 0.175000& 0.022771&    MACSJ0242.6-2132\\    
\tablenotemark{a}& 0.177000& 0.018028&    MACSJ2229.8-2756\\    
\tablenotemark{a}& 0.173000& 0.019105&    MACSJ0947.2+7623\\    
\tablenotemark{a}& 0.189000& 0.025179&    MACSJ1931.8-2635\\    
\tablenotemark{a}& 0.159000& 0.017263&    MACSJ1532.9+3021\\    
\tablenotemark{a}& 0.159000& 0.023548&    MACSJ1720.3+3536\\    
\tablenotemark{a}& 0.177000& 0.017000&    MACSJ0429.6-0253\\    
\tablenotemark{a}& 0.155000& 0.018828&    MACSJ0329.7-0212\\   
0.451000& 0.137000& 0.009000&    RXJ1347.5-1145\\      
0.460500& 0.129000& 0.018668&    3C295\\               
\tablenotemark{a}& 0.156000& 0.033593&    MACSJ1621.6+3810\\    
\tablenotemark{a}& 0.094000& 0.024668&    MACSJ1311.0-0311\\    
\tablenotemark{a}& 0.135000& 0.010512&    MACSJ1423.8+2404\\    
\tablenotemark{a}& 0.155000& 0.018439&    MACSJ0744.9+3927\\    
0.782000& 0.100000& 0.016140&    MS1137.5+6625\\       
0.892000& 0.114000& 0.020881&    ClJ1226.9+3332\\      
\enddata
\tablecomments{Data taken from \cite{Allen2004}.} 
\tablenotetext{a}{MACS redshifts to be published by H. Ebeling (MNRAS).}
\end{deluxetable}


\begin{thebibliography}{9}

\bibitem[Allen et al.(2002)]{Allen2002} Allen S.W., Schmidt R.W., Fabian A.C.: \textit{Cosmological constraints from the X-ray gas mass fraction in relaxed lensing clusters observed with Chandra.} Mon. Not. R. Astron. Soc. \textbf{334} L11-L15 (2002)

\bibitem[Allen et al.(2003)]{Allen2003} Allen S.W., et al.: \textit{Cosmological constraints from the local X-ray luminosity function of the most X-ray-luminous galaxy clusters.} Mon. Not. R. Astron. Soc. \textbf{342} 287-298 (2003)

\bibitem[Allen et al.(2004)]{Allen2004} Allen S.W., et al.: \textit{Constraints on dark energy from Chandra observations of the largest relaxed cluster.} To appear in Mon. Not. R. Astron. Soc. (2004)  Los Alamos e-Print Archive \texttt{astro-ph/0405340}

\bibitem[Armendariz-Picon et al.(2000)]{Armendariz} Armendariz-Picon C., et al.: \textit{Dynamical solution to the problem of a small cosmological constant and late-time cosmic acceleration.} Phys. Rev. Lett. \textbf{85} 4438 (2000)

\bibitem[Babuorova \& Frolov(2003)]{Babourova} Babourova O.V., Frolov B.N.: \textit{Matter with dilaton charge in Weyl-Cartan spacetime and evolution of the universe.} Class. Quantum Grav. \textbf{20} (2003) 1423-1442

\bibitem[Barris et al.(2003)]{Barris} Barris B.J., et al.: \textit{23 high redshift supernovae from the IfA Deep Survey: Doubling the SN sample at z$\ge$0.7.} Astrophys. J. \textbf{602} 571-594 (2004)

\bibitem[Barrow(1990)]{Chaplygin0} Barrow J.D.: \textit{Graduated inflationary universes.} Phys. Lett. \textbf{B235} 40-43 (1990)

\bibitem[Bento et al.(2002)]{Chaplygin9} Bento M.C., Bertolami O., Sen A.A.: \textit{Generalized Chaplygin gas as a scheme for unifaction of dark energy and dark matter.} Los Alamos e-print Archive \texttt{astro-ph/0210375}

\bibitem[Bialek et al.(2001)]{Bialek} Bialek J.J., Evrard A.E., Mohr J.J.: \textit{Effects of preheating on X-ray scaling relations in galaxy clusters.} Astrophys. J. \textbf{555} 597-612 (2001)

\bibitem[Bili\'{c} et al.(2002)]{Chaplygin5} Bili\'{c} N., Tupper G.B., Jain R.D.: \textit{Unification of dark matter and dark energy.} Phys. Lett. \textbf{B535} 17-21 (2002)
 
\bibitem[Caldwell(2002)]{Caldwell} Caldwell R.: \textit{A phantom menace? Cosmological consequences of a dark energy component with super-negative equation of state.} Phys. Lett. \textbf{B454} 23-29 (2002)

\bibitem[Cen \& Ostriker(1994)]{Cen} Cen R., Ostriker J.P.: \textit{X-ray clusters in a cold dark matter + $\Lambda$ universe: A direct, large-scale, high-resolution, hydrodynamic simulation.} Astrophys. J. \textbf{429} (1994) 4-21

\bibitem[Chaboyer et al.(1998)]{Chaboyer} Chaboyer B., et al.: \textit{The age of globular clusters in the light of HIPPARCOS: Resolving the age problem?} Astrophys. J. \textbf{494} (1998) 96-110

\bibitem[Chiba et al.(2000)]{Chiba} Chiba T., et al.: \textit{Kinetically driven quintessence.} Phys. Rev. \textbf{D62} 023511 (2000)

\bibitem[Daly(1994)]{Daly1994} Daly R.A.: \textit{Cosmology with powerful extended radio sources.} Astrophys. J. \textbf{426} 38-50 (1994)

\bibitem[Daly \&  Guerra(2002)]{Daly2002} Daly R.A, Guerra E.J.: \textit{Quintessence, cosmology, and Fanaroff-Riley type IIb radio galaxies.} Astrophys. J. \textbf{124} 1831-1838 (2002)

\bibitem[Daly \&  Djorgovski(2003)]{Daly2003} Daly R.A, Djorgovski S.G.: \textit{A model-independent determination of the expansion and acceleration rates of the universe as a function of redshift and constraints on dark energy.} Astrophys. J. \textbf{597} 9-20 (2003)

\bibitem[Deffayet et al.(1999)]{Deffayet} Deffayet C., et al.: \textit{Accelerated universe from gravity leaking to extra dimensions.}  Phys. Rev. \textbf{D65} 044023 (1999)

\bibitem[Dev et al.(2002)]{Chaplygin4} Dev A., Alcaniz J.S., Jain D.: \textit{Cosmological consequences of a Chaplygin gas dark energy.} Phys. Rev. \textbf{D67} 023515 (2003)

\bibitem[Eke et al.(1998)]{Eke1998} Eke V.R., Navarro J.F., Frenk C.S.: \textit{The evolution of X-ray clusters in a low density universe.} Astrophys.\ J.\ \textbf{503} (1998) 569-592

\bibitem[Fabris et al.(2002)]{Chaplygin2} Fabris J.C., et al.: \textit{Density perturbations in an universe dominated by the Chaplygin gas.} Gen. Rel. Grav. \textbf{34} 53-63 (2002)

\bibitem[Freedman et al.(2001)]{Freedman} Freedman W., et al.: \textit{Final results from the Hubble Space Telescope Key Project to measure the Hubble constant.} Astrophy. J. \textbf{553} 47-73 (2001)

\bibitem[Freese \& Lewis(2002)]{FreeseLewis} Freese K., Lewis M.: \textit{Cardassian expansion: a model in which the universe is flat, matter dominated, and accelerating.} Phys. Lett. \textbf{B540} (2002) 1-8

\bibitem[Fukugita et al.(1998)]{Fukugita1998} Fukugita M., Hogan C.J., Peebles P.J.E.: \textit{The cosmic baryon budget.} Astrophys. J. \textbf{503} (1998) 518-530

\bibitem[Garnavich et al.(1998)]{Garnavich1} Garnavich P.M., et al.: \textit{Constraints on cosmological models from Hubble space telescope observations of high-z supernovae.} Astrophys. J. \textbf{493} L53-57 (1998)

\bibitem[Gonz\'{a}lez-D\'{\i}az(2002)]{Chaplygin7} Gonz\'{a}lez-D\'{\i}az P.F.: \textit{Unified model of dark energy.} Phys. Lett. \textbf{B562} (2003) 1-8

\bibitem[Guerra et al.(2000)]{Guerra2000} Guerra E.J., Daly R.A., Wan L.: \textit{Global cosmological parameters determined using classical double radio galaxies.} Astrophys. J. \textbf{544} 659-670 (2000)

\bibitem[Hamuy et al.(1996)]{Hamuy1996} Hamuy M., et al.: {\it The absolute luminosities of the Cal\'an/Tololo type Ia supernovae.} Los Alamos e-print Archive \texttt{astro-ph/9609059}

\bibitem[Hassa$\ddot{\rm{\i}}$ne et al.(2001)]{Chaplygin1} Hassa$\ddot{\rm{\i}}$ne M., et al.: \textit{Relativistic Chaplygin gas with field-dependent Poincar\'{e} symmetry.} Lett. Math. Phys. \textbf{57} 33-40 (2001)

\bibitem[Hehl et al.(1995)]{PhysRep}  Hehl F.W., McCrea J.D., Mielke E.W., Ne\'{}eman Y.: \textit{Metric-affine gauge theory of gravity: Field equations, Noether identities, world spinors, and breaking of dilation invariance.} Phys. Rep. \textbf{258} (1995) 1-171

\bibitem[Kamenshchik et al.(2001)]{Chaplygin3} Kamenshchik A., et al.: \textit{An alternative to quintessence.} Phys. Lett. \textbf{B511} 265-268 (2001)

\bibitem[Kamionkowski \& Turner(1990)]{KamionTurn} Kamionkowski M., Turner M.S.: \textit{Thermal relics: Do we know their abundances?} Phys. Rev. \textbf{D42} 3310-3320 (1990)

\bibitem[Khalatnikov \& Kamenshchik(2003)]{Khalatnikov} Khalatnikov I.M., Kamenshchik A.Y.: \textit{A generalisation of the Heckmann - Schuecking cosmological solution.} Phys. Lett. \textbf{B553} 119-125 (2003) 

\bibitem[Kirkman et al.(2003)]{Kirkman2003} Kirkman D., et al.: \textit{The cosmological baryon density from the deuterium-to-hydrogen ratio in QSO absorption systems: D/H toward Q1243+3047.} Astrophy. J. Supp. \textbf{149} 1-28 (2003)

\bibitem[Kolb \& Turner(1990)]{KolbTurner}  Kolb E.W, Turner M.S.: \textit{The early universe.} Addison-Wesley, Redwood City (1990)

\bibitem[Obukhov \& Tresguerres(1993)]{Obukhov2} Obukhov Y.N., Tresguerres R.: \textit{Hyperfluid - a model of classical matter with hypermomentum.} Phys. Lett. \textbf{A184} 17-22 (1993)

\bibitem[Obukhov et al.(1997)]{Obukhov} Obukhov Y.N., Vlachynsky E.J., Esser W., Hehl F.W.: \textit{Effective Einstein theory from metric-affine gravity models via irreducible decompositions.} Phys. Rev. \textbf{D56} 12 (1997) 7769-7778

\bibitem[O'Meara et al.(2001)]{Meara} O'Meara J., et al.: \textit{The deuterium to hydrogen abundance ratio toward a fourth QSO: HS 0105+1619.} Astrophy. J. \textbf{552} 718-730 (2001)

\bibitem[\"Ozer \& Taha(1987)]{Ozer} \"Ozer M., Taha M.O.: \textit{A model of the universe free of cosmological problems.} Nuc. Phys. \textbf{B287} 776-796 (1987)

\bibitem[Padmanabhan(2002)]{PadmanabhanAstro3} Padmanabhan T.: \textit{Theoretical astrophysics Volume III: Galaxies and Cosmology.} Cambridge University Press, Cambridge (2002)

\bibitem[Pen(1997)]{Pen1997} Pen U.: \textit{Measuring the universal deceleration using angular diameter distances to clusters of galaxies.} New Astronomy \textbf{2} (1997) 309-317

\bibitem[Perlmutter et al.(1997)]{Perlmutter2} Perlmutter S., et al.: \textit{Measurements of the cosmological parameters $\Omega$ and $\Lambda$ from the first seven supernovae at $z \ge 0.35$.} Astrophys. J. \textbf{483} (1997) 565-581

\bibitem[Perlmutter et al.(1999)]{Perlmutter} Perlmutter S., et al.: \textit{Measurements of $\Omega$ and $\Lambda$ from 42 high-redshift supernovae.} Astrophys.\ J.\ \textbf{517} (1999) 565

\bibitem[Puetzfeld(2002a)]{Puetzfeld1}  Puetzfeld D.: \textit{A cosmological model in Weyl-Cartan spacetime: I. Field equations and solutions.} Class. Quantum Grav. \textbf{19} (2002) 3363-3280  

\bibitem[Puetzfeld(2002b)]{Puetzfeld2}  Puetzfeld D.: \textit{A cosmological model in Weyl-Cartan spacetime: II. Magnitude-redshift relation.} Class. Quantum Grav. \textbf{19} (2002)  

\bibitem[Puetzfeld(2004)]{Puetzfeld2004} Puetzfeld D.: \textit{Status of non-Riemannian cosmology.} Proc. 6th UCLA Symp. on "Sources and Detection of Dark Matter and Dark Energy in the Universe", February 18-20, 2004, to be published, Los Alamos e-Print Archive \texttt{gr-qc/04041194}

\bibitem[Puetzfeld \& Chen(2004)]{PuetzfeldChen} Puetzfeld D., Chen X.: \textit{Testing non-standard cosmological models with supernovae.} Class. Quantum Grav. \textbf{21} (2004) 2703-2722

\bibitem[Randall \& Sundrum(1999a)]{Randall1} Randall L., Sundrum R.: \textit{Large Mass Hierarchy from a Small Extra Dimension.} Phys. Rev. Lett. \textbf{83} 3370-3373 (1999)

\bibitem[Randall \& Sundrum(1999b)]{Randall2} Randall L., Sundrum R.: \textit{An Alternative to Compactification.} Phys. Rev. Lett. \textbf{83} 4690-4693 (1999)

\bibitem[Ratra \& Peebles(1988)]{Ratra} Ratra B., Peebles P.J.E.: \textit{Cosmological consequences of a rolling homogeneous scalar field.}  Phys. Rev. \textbf{D37} 3406 (1988)

\bibitem[Riess et al.(1998)]{Riess2} Riess A.G., et al.: \textit{Observational evidence from supernovae for an accelerating universe and a cosmological constant.} Astrophys. J. \textbf{116} (1998) 1009-1038

\bibitem[Riess et al.(2001)]{Riess1} Riess A.G., et al.: \textit{The farthest known supernova: Support for an accelerating universe and a glimpse of the epoch of deceleration.} Astrophys. J. \textbf{560} (2001) 49-71

\bibitem[Riess et al.(2004)]{Riess2004} Riess A.G., et al.: \textit{Type Ia supernovae discoveries at $z>1$ form the Hubble Space Telescope: Evidence for past deceleration and contraints on dark energy evolution.}  Astrophys. J. \textbf{607} (2004) 665-887

\bibitem[Sasaki(1996)]{Sasaki1996} Sasaki S.: \textit{A new method to estimate cosmological parameters using the baryon fraction of clusters of galaxies.} Pub. Astron. Soc. Japan \textbf{48} (1996) L119-L122

\bibitem[Schmidt et al.(1998)]{Schmidt} Schmidt B.P., et al.: \textit{The high-z supernova search: Measuring cosmic deceleration and global curvature of the universe using type Ia supernovae.} Astrophys. J. \textbf{507} (1998) 46-63

\bibitem[Spergel et al.(2003)]{Spergel2003} Spergel D.N., et al.: \textit{First year Wilkinson microwave anisotropy probe (WMAP) observations: Determination of cosmological parameters.}  Astrophys. J. Suppl. \textbf{148} (2003) 175-194 

\bibitem[Tonry et al.(2003)]{Tonry} Tonry J.L., et al.: \textit{Cosmological results from high-z supernovae.}  Astrophys. J. \textbf{594} (2003) 1-24

\bibitem[Truran et al.(2001)]{Truran} Truran J.W., et al.: \textit{Nucleosynthesis clocks and the age of the galaxy.} ASP Conference Series, Vol. TBD, 2001, Eds. T. von Hippel, N. Manset, C. Simpson Los Alamos e-print Archive \texttt{astro-ph/0109526}

\bibitem[Vishwakarma(2001)]{Vishwakarma3} Vishwakarma R.G.: \textit{Consequences on variable $\Lambda$-models from distant type Ia supernovae and compact radio sources.} Class. Quantum Grav. \textbf{18} 1159-1172 (2001)

\bibitem[Wang(2000b)]{Wang} Wang Y.: \textit{Flux-averaging analysis of type Ia supernova data.} Astrophys. J. \textbf{536} 531-539 (2000)

\bibitem[Wetterich(1988)]{Wetterich} Wetterich C.: \textit{Cosmology and the fate of dilatation symmetry.} Nuc. Phys. \textbf{B302} 668-696 (1988)

\bibitem[Zhu \& Fujimoto(2002)]{Zhu2002} Zhu Z.H., Fujimoto M.K.: \textit{Cardassian expansion: Constraints from compact radio source angular size versus redshift data.} Astrophys. J. \textbf{581} 1-4 (2002)
     
\bibitem[Zhu \& Fujimoto(2003)]{Zhu2003} Zhu Z.H., Fujimoto M.K.: \textit{Constraints on Cardassian expansion from distant type Ia supernovae.} Astrophys. J. \textbf{585} 52-56 (2003)

\bibitem[Zhu et al.(2004)]{Zhu1} Zhu Z.H., Fujimoto M.K., He X.T.: \textit{Observational constraints on cosmology from modified Friedmann equations.} Astro. Phys. Journ. \textbf{603} 365-370 (2004)

\end{thebibliography}
\end{document}